\documentclass[twocolumn,aps,prb,showkeys,groupaddress,superscriptaddress,sectionbib,square]{revtex4-1}
\setcitestyle{numbers,square}
\usepackage{multirow}
\usepackage{siunitx}  

\providecommand{\tabularnewline}{\\}

\usepackage{epsfig}
\usepackage{blindtext}

\newcommand{\angstrom}{\mbox{\normalfont\AA}}

\begin{document}

\title{Half-metallic compositional ranges for selected Heusler alloys}

\author{Nikolai~A. Zarkevich}\email{zarkev@ameslab.gov}

\affiliation{Ames Laboratory, U.S. Department of Energy, Ames, Iowa 50011 USA}

\author{Prashant Singh}

\affiliation{Ames Laboratory, U.S. Department of Energy, Ames, Iowa 50011 USA}

\author{A.~V. Smirnov}
\affiliation{Ames Laboratory, U.S. Department of Energy, Ames, Iowa 50011 USA}

\author{Duane~D. Johnson}\email{ddj@ameslab.gov}
\affiliation{Ames Laboratory, U.S. Department of Energy, Ames, Iowa 50011 USA}
\affiliation{Departments of Materials Science \& Engineering, Iowa State University, Ames, Iowa 50011 USA}

\date{\today}

\begin{abstract}
For a material that is a half-metal, there should exist a range of compositions for half-metallicity. 
This compositional range  can be expressed in terms of {\it electron count} and computed. 
We investigate electronic and magnetic properties of doped full- and half-Heusler alloys (stoichiometry XYZ$_{2}$ and XYZ, respectively)
 with elements X from groups 13\textendash16 and periods 3\textendash6 of the Periodic Table, Y=\{Mn, Fe\}, and Z=\{Co, Ni\}.
Using spin density functional theory, we predict shifts of the Fermi energy in the doped and solid-solution alloys. 
These predictions can be used for band-gap engineering of multi\-component half-metals and provide the viable range of compositions, 
such as for a range of $n  \! = \!  x \! + \! y \! + \! z$ in (Co$_{2-z}Ni_z)$(Mn$_{1-y}$Fe$_{y}$)(Sn$_{1-x}$Sb$_{x}$). 
This methodology for doped and chemically disordered half-metallic alloys offers a design approach to electronic-structure engineering
 that can accelerate development of  half-metals for novel electronic and spintronic applications.
\end{abstract}



\maketitle

\section{Introduction}
{\par} Materials discovery for alloys with specific properties, particularly for viable concentration ranges for half-metallicity,  is greatly accelerated by theoretical guidance \cite{JPhysD51n2p024002y2018,NRM3n5p5y2018,NRM2p17053y2017,NRM1p15004y2016,NMat14n10p973y2015,JMS47n21p7317y2012,ChemSci11n10p2696y2020,npjCompMat5n1p41y2020},
especially in the electronic-structure engineering of doped or chemically disordered alloys.
The associated compositional (and electron count) change alters electronic, magnetic, and other (including surface  \cite{Kawasaki2018}) physical properties \cite{Galanakis2002_1,Hanssen1990,Complexity11p36y2006,Raphael2002}.
Here we study the electronic structure and its dependence on composition  in fully and partially-ordered Heusler alloys, some of which are half-metallic.
Our results provide valuable guidance for adjusting the width and position of band gaps in the minority spins relative to the Fermi energy, $E_{F}$.

{\par} Heusler alloys \cite{Heusler1903} are a large class of naturally occurring \cite{JSSChem43p354y1982} and manufactured \cite{HeuslerAlloys2016} ternary intermetallic compounds
with L2$_{1}$ full Heusler (fH) or C1$_{b}$ half-Heusler (hH) structure,  
which can be partially disordered \cite{SciRep8n1p9147y2018,ActaMat142p49y2018,ActaMat148p216y2018,ActaMat115p308y2016,ActaMat107p1359y2016,ActaMat104p210y2016,PRB94p024418y2016,JETPlett107p107y2018,PhysRevMaterials3p114406y2019}. 
The known competing structures are the inverse Heusler and hexagonal $D0_{19}$ \cite{JMMM501p166426y2020,JAC771p793y2019}. 
Many Heuslers are ferromagnetic (FM), with Curie temperatures $T_c$ between
 200 and 1100 K \cite{Buschow1983,PJW1971,Brown2000}. 
The ones with $T_c$ near room temperature were considered for magnetocalorics \cite{JPhysD51n2p024002y2018,JPhysD49n39p395001y2016}. 
Some alloys have  a band gap in the minority spins, which may (as in half-metals 
\cite{Groot1983,Hanssen1986,JPhysCM1p2341y1989,Fujii1989,Fujii1990,Fuji1995,Galanakis2002,Galanakis2002_1,Galanakis2006,RSCadv9p30462y2019,PRMaterials3p024410y2019}) 
or may not be at $E_{F}$.  

{\par} A half-metal conducts electrons with one spin orientation, but acts as an insulator or semiconductor 
(with a band gap at $E_{F}$) for electrons of opposite spins.
Half-metals are  used in small-scale magnetic and spin filtering devices \cite{Galanakis2005,AdvMaterials32n6p1905734y2020,Lukyanenko2019}
and switches  \cite{NatureComm5p3682y2014}, and find applications in magnetic materials \cite{Krishnan2016}. 
The band-gap engineering \cite{Science235n4785p172y1987,Science352n6292p1446y2016}
helps to create materials for advanced electronic devices. 
Such solid-state devices are used in spintronics -- a rapidly developing branch of electronics, incorporating
electron spin degree of freedom (DoF) \cite{Science294p1488y2001,Spintronics2013}.

{\par} We focus on the electronic properties of doped fH\textendash{Z$_{2}$MnX} and hH\textendash{ZMnX} alloys,
 with Z=\{Co, Ni\} and X=\{Al, Ga, In, Tl; Si, Ge, Sn, Pb; P, As, Sb, Bi; S, Se, Te\}, elements from
groups 13\textendash 16 and periods 3\textendash 6 in the Periodic Table \cite{MendeleevTable}. 
We compute the electronic structure of these alloys and predict compositional intervals for half-metalicity in 
terms of electron count $n$. 
We extend compositions to address Mn alloyed with Fe in Co$_2$(Mn$_{1-y}$Fe$_{y}$)A, 
where A is Sn or Sb, and $0 \! \le \! y \! \le \! 1$. 

{\par} We find that in  fH\textendash{Co$_{2}$MnX} the width of the minority-spin band gap $E_{gap}$ 
decreases with period (3\textendash 6) and increases with group (13\textendash 16) of X. 
Most of these alloys are stable. 
Due to small mixing enthalpies, they allow substitutional disorder on each sub\-lattice at room temperature $T$.
For the  hH\textendash{NiMnX} alloys considered, the gap in the minority-spin manifold lies at or slightly above $E_{F}$. 
Five of them (with X=Si, Ge; P, As, Sb) could be half-metals, but only NiMnSb is stable. 
The structural stability restricts design of hH half-metals.

{\par}Often half-metals have lower formation energies (hence are more stable) than their metal counterpart. 
 Interestingly,  we predict a possible existence of weakly stable multicomponent half-metallic bismuthides. 
Considering a doped hH\textendash{NiMnBi}, we predict that adding from 0.3 to 1.3 electrons per formula unit ($e^{-}$/f.u.)
would turn this compound into a half-metal.
We confirm this prediction for the quaternary CuNi$_{3}$Mn$_{4}$Bi$_{4}$ and ZnNi$_{7}$Mn$_{8}$Bi$_{8}$ alloys. 
We analyze   stability  by considering relative structural energies and cross-sections of the ground-state formation energies versus composition (sometime referred to as the ground-state hull).

{\par}  This paper is organized by Section~\ref{CrystalStructure}  specifying the crystal structures and compositional ranges,
Section~\ref{sec:Calculational-synopsis} identifying computational methods, Section \ref{Results} containing results and discussion,
 followed by a summary in Section~\ref{Summary}. 
Technical details are provided in Appendix~\ref{Appendix}. Metallic fH--Ni$_2$MnX  is considered in Appendix~\ref{Ni2MnX}.

\section{\label{CrystalStructure}Structures and Compositions}

\begin{figure}[t]
\includegraphics[width=45mm]{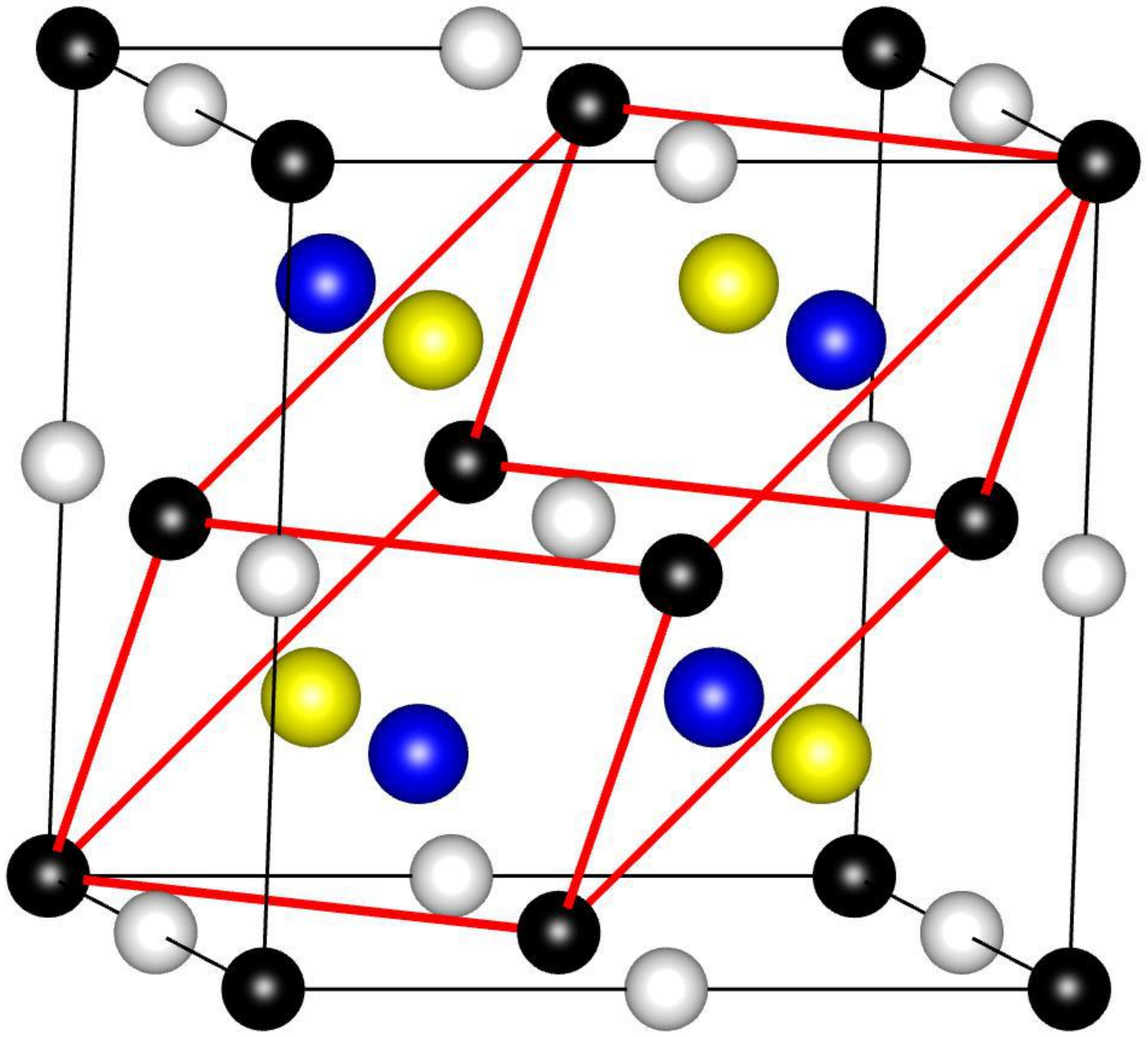} 
\caption{\label{fig1str}  Heusler MM'YX structure [$F\bar{4}3m$, space group \#216] 
has atoms on bcc sites, denoted as M (black), M' (white), Y=\{Mn, Fe\} (blue), and X (yellow). 
Conventional (black) and primitive (red) unit cells are outlined.
C1$_{b}$, cF12  hH-MYX has vacant M' $(\frac{1}{2}$,$\frac{1}{2}$,$\frac{1}{2})$ sites.  
L2$_{1}$, cF16 ($Fm\bar{3}m$, \#225) fH-Z$_{2}$YX has Z on both M and M' sites.  
Inverse Heusler has the same atoms on M and X sites, but M and M' differ.}
\end{figure}

{\it {\label{Structure}}Crystal Structure:} The full-Heusler and half-Heusler alloys crystallize in  L2$_{1}$ and C1$_{b}$ structures, respectively. 
Fig.~\ref{fig1str} represents the quaternary LiMgPdSn-type Heusler,  Cu$_2$MnAl-type fH,  AgAsMg-type hH, 
and the inverse-Heusler structures;  caption provides the space group, etc. 
The fH-Z$_{2}$YX (L2$_{1}$, cF16) structure with Cu$_{2}$MnAl prototype has a 16-site conventional cubic unit cell with 3 atom types (decorating a body-centered cubic (bcc) lattice), consisting of four interpenetrating fcc sub\-lattices with positions (0,0,0) and ($\frac{1}{2}$,$\frac{1}{2}$,$\frac{1}{2}$) for Z, ($\frac{1}{4}$,$\frac{1}{4}$,$\frac{1}{4}$) for Y (Mn) and ($\frac{3}{4}$,$\frac{3}{4}$,$\frac{3}{4}$)  for X atoms. The primitive unit cell has only 4 sites (outlined in red in Fig.~\ref{fig1str}).  The hH-M$_{1}$YX (C1$_{b}$, cF12) structure with MgAgAs prototype differs from cF16 fH by a vacancy on ($\frac{1}{2}$,$\frac{1}{2}$,$\frac{1}{2}$) sites. 

{\it {\label{Composition}} Compositions:} We consider fH-Z$_{2}$YX and hH-M$_{1}$YX systems with
Z=\{Co, Ni\};  M=\{Ni\}; Y=\{Mn\}; and X=\{Al, Ga, In, Tl; Si, Ge, Sn, Pb; P, As, Sb, Bi; S, Se, Te\}, 
as well as solid-solution Co$_2$(Mn$_{1-x}$Fe$_{x}$)A with A=\{Sn or Sb\}.  
Our investigation covers both fully-ordered and partially-disordered (substitutional) alloys.

\section{\label{sec:Calculational-synopsis} Synopsis of Methods}
\subsection{\label{Guidance}Guidance for materials engineering} 
{\it {Electron count \& composition}:}~Historically, the search for half-metals
started from line compounds \cite{Buschow1983,Irkhin1994,NatComm5p3974y2014,JETP118p426y2014}. 
We emphasize that $E_{F}$ remains in the band gap within a range of energies, 
bounded by the highest occupied and the lowest unoccupied states in the minority spins, 
and there is a corresponding range of half-metallic compositions. 

\begin{figure}[t]
\includegraphics[width=60mm]{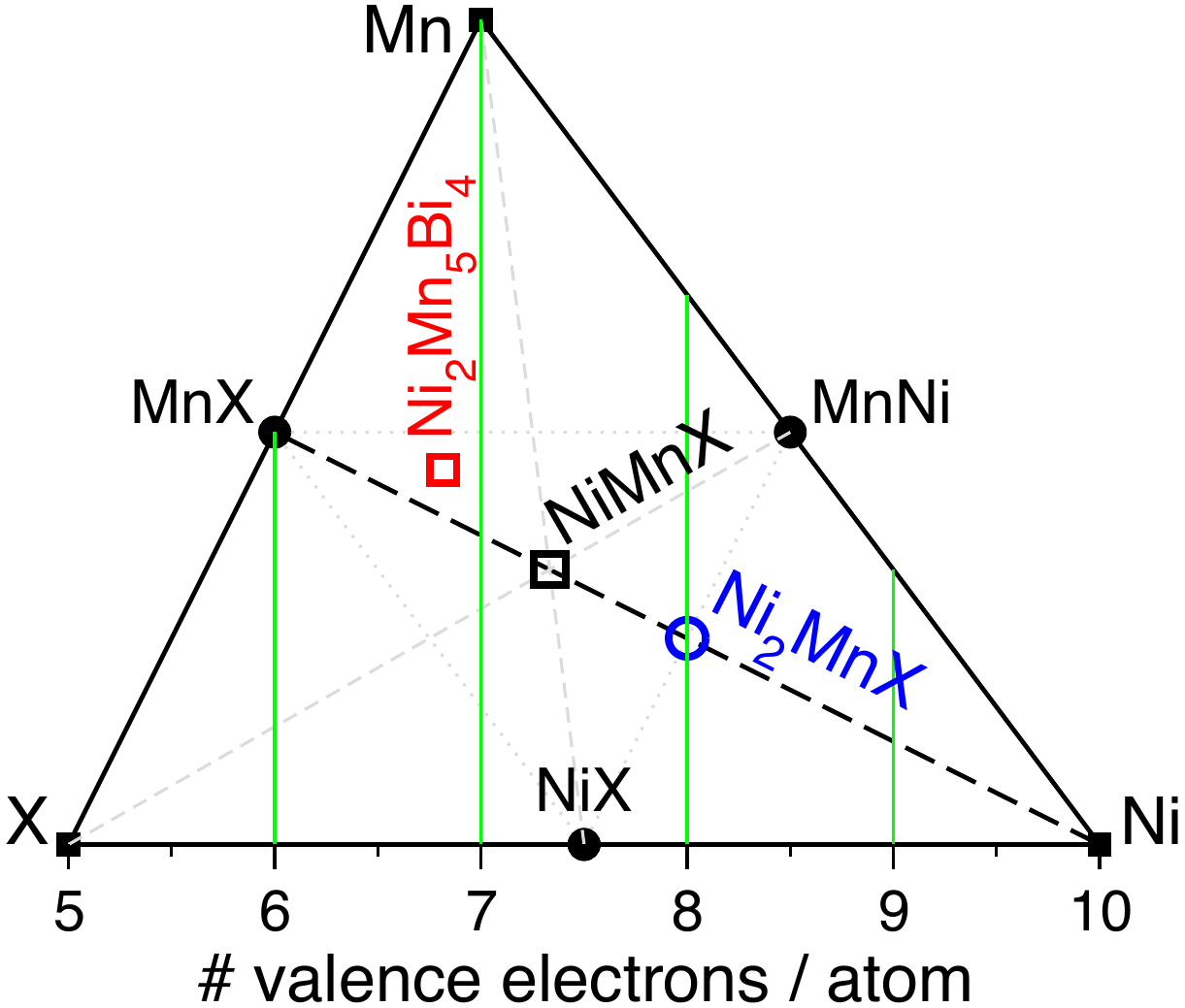} 
\caption{\label{fig_el} Isosurfaces of electron count per atom (vertical lines) for Ni-Mn-X with elements X = \{P, As, Sb, Bi\}. }
\end{figure}

{\par} Indeed, the number of valence electrons $N (c)$ depends on composition $c$.  
In Figure~\ref{fig_el} we illustrate the inverse [many-to-one $c(N)$] relation for the ternary Ni-Mn-X alloys. 
A fixed $N$ specifies a set of compositions,  leaving sufficient freedom in choosing elements for doping. 
For example,  hH alloys with compositions NiMn(P$_x$As$_y$Sb$_z$Bi$_w$), 
where $0 \le \{x,y,z,w \} \le 1$ and $x+y+z+w=1$, have $N \! = \! 7\frac{1}{3}$, see Fig.~\ref{fig_el}.
Below we express the electron count $n$ as a difference in the number of valence electrons relative to a line compound. 

{\it {\label{GuidanceRange}}Half-metallic range:}~In each compound,  we find a band gap in the minority spins (if any)
of the width $E_{gap} = (E_+ - E_-)$ at the electronic energies $E$ in the range $E_- \! \le \! ( E - \text{E}_F ) \! \le \! E_+$. 
A compound is  half-metallic, if $E_{F}$ happens to be in this gap (thus, $E_- \! < \! 0$ and $E_+ \! > \! 0$). 

{\par} The number of electronic states below $E_0$ is
\begin{equation}
\label{eqN}
N (E_0) = \int_{-\infty}^{E_0} dE~ g(E ) ,
\end{equation}
where $g \ge 0$ is the electronic density of states (DOS) for both spins.
At zero electronic temperature, all states below the $E_{F}$ are filled, hence $N(E_F)$ is equal to the total number of electrons $N$.  
(If one uses pseudo\-potentials or does not count core electrons, then one can use the total number of valence electrons.)
A band gap in the minority spins covers a range of both occupied and empty states from $ N_- \equiv N(E_F + E_-)$ to $ N_+ \equiv N(E_F + E_+)$;   those are the edges of the band gap.   Next, we use eq.~\ref{eqN} to find the differences 
\begin{equation}
 \label{eqndef}
 n_{\pm} \equiv (N_{\pm} -N) = \int_{E_F}^{E_F + E_\pm} g(E ) dE .
\end{equation}
A compound is  half-metallic, if $n_- <0$ and $n_+ >0$, see Tables~\ref{t2n} and \ref{t4n}.

{\par} Within the frozen-band approximation, let us increase the total number of electrons from $N$ to $(N+n)$;   
negative $n$ stands for subtracting electrons. 
The added or subtracted electrons will shift  $E_{F}$ to the band gap (thus making the system half-metallic), if 
\begin{equation}
 \label{eqn}
  n_- \le n \le  n_+ .
\end{equation}

\begin{figure*}[t]
\centering{
\includegraphics[width=84mm]{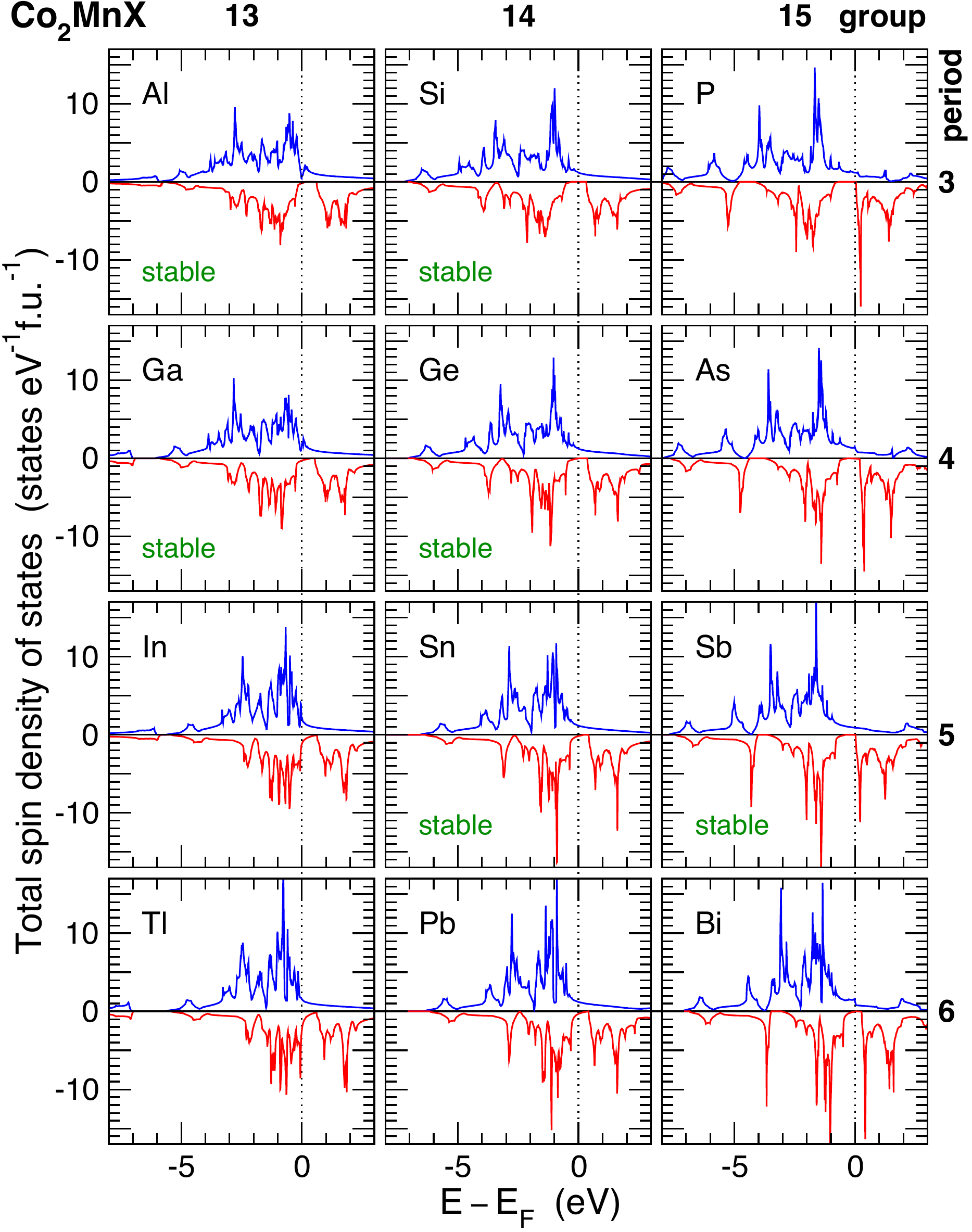}  \hspace{3mm}
\includegraphics[width=84mm]{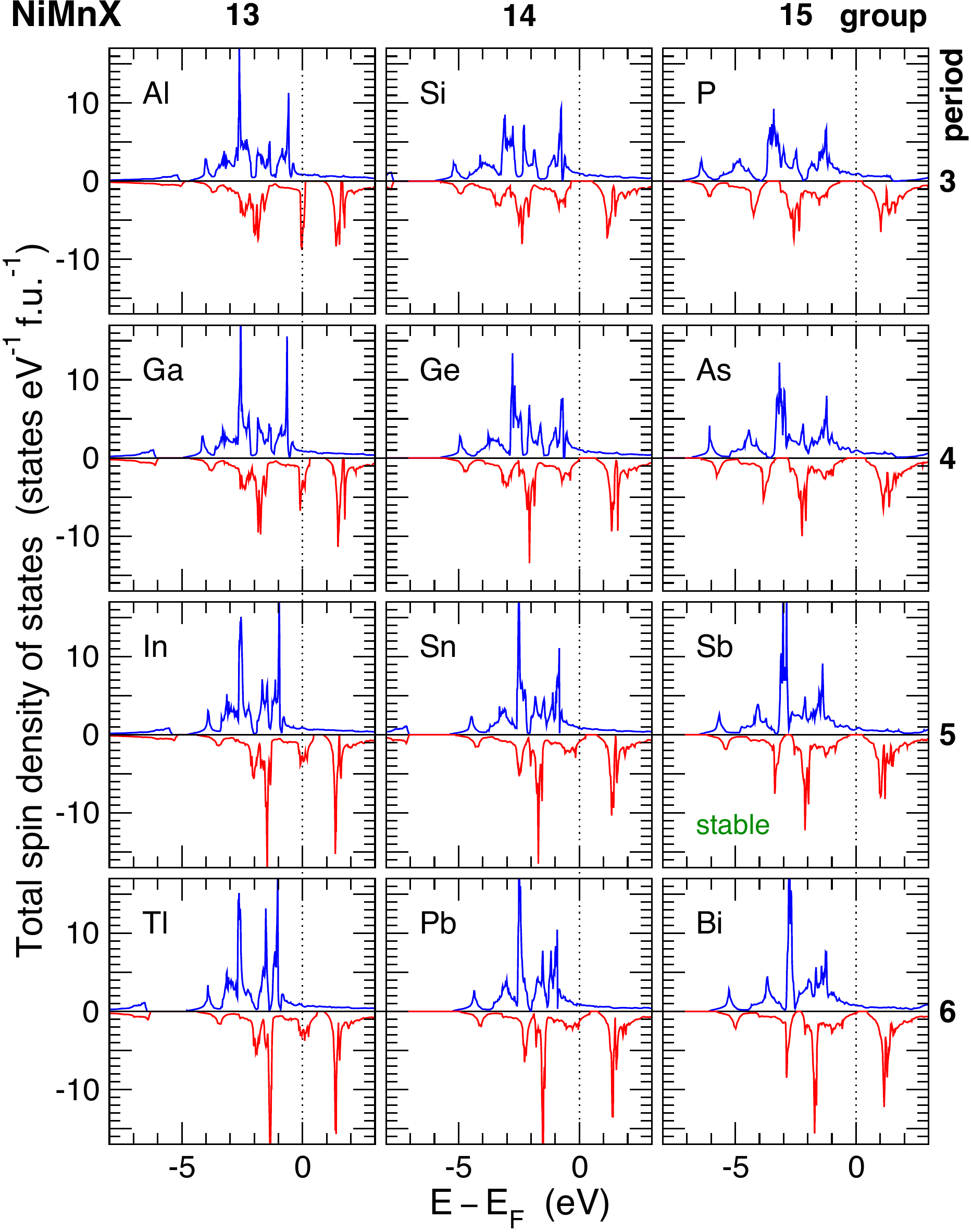}
}
\caption{\label{fig5dos} 
Electronic spin DOS (states/eV$\cdot$f.u.) for fH-Co$_{2}$MnX (left) and hH-NiMnX (right)  
for 12 elements X (group 13-15, period 3-6). Stable compounds are known from experiment \cite{Buschow1983,PJW1971,JAP105n1p013716y2009}. }
\end{figure*}

{\par}  In general, doping a conductor with electronic donors or acceptors changes the number of electrons, shifts $E_{F}$, and adds impurity states.  If those impurity states are not in the band gap, then half-metallicity of the doped system is approximated by inequality (\ref{eqn}). 

{\par}   Lattice relaxations around impurities and defects broaden the bands and narrow the band gap. 
In addition, sometimes electronic bands of the dopant can appear in the band gap. 
These effects narrow both the band gap and the half-metallic compositional range. 
With this caution, inequality (\ref{eqn}) and  Tables~\ref{t2n} and  \ref{t4n} 
provide a practical guidance for engineering of the advanced half-metals. 

\subsection{\label{Methods}Computational Methods}

{\it DFT codes:}~We calculate energetics and electronic structure using density functional theory (DFT). 
To verify accuracy of the predicted structural energies and electronic bands in ordered and disordered alloys, 
we use three DFT codes: a pseudo-potential plane-wave code implemented in VASP \cite{VASP1,VASP2}, 
an all-electron KKR-CPA Green's Function code MECCA \cite{MECCA}, and a TB-LMTO-vLB code with a locally corrected exchange potential \cite{PRB93p085204y2016} that yields bandgaps 
comparable to hybrid exact-exchange and GW methods \cite{datta2019,PhysRevB.96.054203,Carbon2020}, but with the speed of local density approximation,  see Appendix \ref{Appendix} for details ($k$-mesh grids, exchange-correlation, etc.). 

{\it Structural stability:}~To address stability, we calculate for each structure the formation energies  versus composition. 
Each structure on the ground-state (GS) hull is stable and has a nonpositive (negative or zero) formation energy. 
Structures with energies above the GS hull are either unstable or meta\-stable.
They might be stabilized by entropy at finite temperature $T$; they can transform or segregate with energy release at lower $T$.  

{\it Accuracy of bandgap widths:}~The calculated band gap in each spin channel depends on the exchange-correlation functional. 
Both local density approximation (LDA) and generalized gradient approximation (GGA) usually underestimate the band gap in semiconductors
and half-metals;
this systematic error narrows the predicted compositional range for half-metals. 
An error in the predicted band gaps is larger in LDA, smaller in GGA \cite{GGA}, 
and is expected to cancel for the exact exchange and correlation. 
Hence, we check results in Tables~\ref{t2n} and \ref{t4n} using a spin-polarized version of van Leeuwen and Baerends (vLB) correction to LDA; 
with speed of local functionals, 
 this LDA+vLB approximation for exact exchange reproduces hybrid-exchange bandgaps \cite{PRB93p085204y2016,datta2019,PhysRevB.96.054203}.

\begin{table}[b]
\caption{\label{t2n}  
The minority-spin band gap width $E_{gap}$ (eV) and position in terms of energies $E_{\pm}$ (relative to $E_F$) and electron count $n_{\pm}$ 
 in full-Heusler Co$_{2}$MnX ternary line compounds.  
 The unstable compounds are marked with an asterisk ($^{*}$). 
}
\begin{tabular}{|l|rl|rl|ll|}
\hline 
\multirow{2}{*}{ X} & $n_{-}$  & $n_{+}$  & $E_{-}$  & $E_{+}$  & $E_{gap}^{GGA}$  & $E_{gap}^{vLB}$ \tabularnewline
 & \multicolumn{2}{c|}{($e$/f.u.)} & \multicolumn{2}{c|}{(eV)} &  \multicolumn{2}{c|}{(eV)} 
 \tabularnewline
\hline 
Al  & 0.07  & 0.84  & 0.06  & 0.62  & 0.56  & 0.38 \tabularnewline
Ga  & 0.5  & 0.84  & 0.24  & 0.55  & 0.31  & 0.23 \tabularnewline
In$^{*}$  & 0.9  & 1.0  & 0.54  & 0.64  & 0.10  & \textendash{} \tabularnewline
Tl$^{*}$  & 1.1  & 1.1  & 0.63  & 0.63  & 0.0  & 0.0 \tabularnewline
\hline 
Si  & -0.6  & 0.35  & -0.39  & 0.34  & 0.73 & 0.49 \tabularnewline
Ge  & -0.2  & 0.4  & -0.12  & 0.38  & 0.50  & 0.37 \tabularnewline
Sn  & 0.1  & 0.45  & 0.09  & 0.44  & 0.35  & 0.37 \tabularnewline
Pb$^{*}$  & 0.35  & 0.5  & 0.28  & 0.43  & 0.15  & 0.11 \tabularnewline
\hline 
P$^{*}$  & -1.1  & 0.05  & -0.76  & 0.04  & 0.80  & 0.55 \tabularnewline
As$^{*}$  & -0.6  & 0.05  & -0.43  & 0.05  & 0.48  & 0.44 \tabularnewline
Sb  & -0.5  & 0.03  & -0.51  & 0.02  & 0.53 & 0.38 \tabularnewline
Bi$^{*}$  & -0.1  & 0.2  & -0.08  & 0.27  & 0.35  & 0.26 \tabularnewline
\hline 
\end{tabular}
\end{table}
\begin{table}[b]
\caption{\label{t4n} 
The minority-spin band gap $E_{gap}$ and its position in terms of energies $E_{\pm}$ and electron count $n_{\pm}$ 
in half-Heusler NiMnX compounds.  See sections~\ref{GuidanceRange} and \ref{eGap}.  
}
\begin{tabular}{|l|rl|rl|ll|}
\hline 
\multirow{2}{*}{ X}  & $n_{-}$  & $n_{+}$  & $E_{-}$  & $E_{+}$  & $E_{gap}^{GGA}$  & $E_{gap}^{vLB}$ \tabularnewline
 & \multicolumn{2}{c}{($e$/f.u.)} &  \multicolumn{2}{c|}{(eV)} &  \multicolumn{2}{c|}{(eV)}   
 \tabularnewline
\hline 
Al$^{*}$  & 0.6  & 1.2  & 0.10  & 0.90 & 0.80  & $-$ \tabularnewline
Ga$^{*}$  & 0.9  & 1.3  & 0.30  & 1.00 & 0.70  & $-$ \tabularnewline
In$^{*}$  & 1.25  & 1.5  & 0.46  & 0.88  & 0.40  & 0.32 \tabularnewline
Tl$^{*}$  & 1.4  & 1.6  & 0.60  & 0.90  & 0.30  & 0.27 \tabularnewline
\hline 
Si$^{*}$  & -0.37  & 0.45  & -0.35  & 0.56  & 0.91  & 0.8 \tabularnewline
Ge$^{*}$  & -0.0  & 0.57  & -0.0  & 0.7  & 0.70 & 0.6 \tabularnewline
Sn$^{*}$ & 0.40  & 0.72  & 0.28  & 0.74  & 0.46  & 0.43 \tabularnewline
Pb$^{*}$  & 0.65  & 0.87  & 0.45  & 0.79  & 0.34  & 0.28 \tabularnewline
\hline 
P$^{*}$  & -0.67  & 0.22  & -0.66  & 0.28  & 0.93  & 0.7 \tabularnewline
As$^{*}$  & -0.37  & 0.30  & -0.38  & 0.39  & 0.77  & 0.6 \tabularnewline
Sb  & -0.17  & 0.22  & -0.27  & 0.28  & 0.55  & 0.45 \tabularnewline
Bi$^{*}$  & 0.06  & 0.31  & 0.07  & 0.47  & 0.40  & 0.34 \tabularnewline
\hline 
\end{tabular}
\end{table}

\section{\label{Results}Results and Discussion}

Known half-metals among ternary systems include full Heusler Co$_{2}$MnX
with X=\{Si, Ge, Sb\} \cite{JAC781p216y2019,PRApplied11p064009y2019} and half-Heusler NiMnSb compounds \cite{EuroPhysJB16n2p287y2020}. 
However, spin polarization depends on composition, ordering \cite{JAC101n2p023901y2007,JMMM502p166536y2020}, and temperature \cite{PRB94p094428y2016}.
We emphasize that not only the line compounds, but also ranges of compositions (with off-stoichiometric disorder) can be half-metallic.
We compute the compositional ranges that keep known half-metals
half-metallic, and predict the level of doping that can turn others
(like Co$_{2}$MnSn and hypothetical NiMnBi) into half-metals.

\subsection{\label{EEE}Data for electronic-structure engineering} 

\noindent {\it {\label{eDOS}} Electronic DOS:} We compare the calculated electronic DOS for the ternary line compounds Co$_{2}$MnX and Ni$_{1}$MnX in Fig.~\ref{fig5dos}. We use experimental data to mark the known stable compounds. The values of $E_{\pm}$ and $n_{\pm}$, defined in section IIIA, are provided in Tables I and II for fH-Co$_{2}$MnX and hH-NiMnX (which have the minority-spin band gap in the vicinity of $E_{F}$). This data is useful for engineering materials with a modified electronic structure and for predicting the band gaps in disordered alloys, some of which will be considered in section~\ref{Disorder}. 

\noindent {\it {\label{eGap}} Minority-spin band gaps:} Interestingly, we find minority-spin band gaps (or pseudo-gaps) in all ferromagnetic compounds (both stable and unstable), see Fig.~\ref{fig5dos}. 
In fH Co$_2$MnX the gaps are at or near $E_F$. Similarly, in hH NiMnX they are at or slightly above $E_F$. 
 In contrast, there are no half-metals among fH Ni$_2$MnX alloys (relegated to Appendix \ref{Ni2MnX}, see Fig.~\ref{DOS_Ni2MnX_fH}). 
Because Ni has more electrons than Co, the difference between the fH Co$_2$MnX and Ni$_2$MnX is expected from the electron count, explained in section~\ref{Guidance}. Indeed, the gaps are near $E_F$ in Co$_2$MnX, but quite far ($\sim \! 1\,$eV) below $E_F$ in Ni$_2$MnX. 

{\par}Tables~\ref{t2n} and \ref{t4n} show the  band gap $E_{gap} = (E_+ - E_-)$ for the minority spins  from GGA (VASP) and LDA+vLB (TB-LMTO) results. The gap extends from $E_-$ to $E_+$ (both energies are relative to $E_{F}$); for half-metals, these are energies of the highest occupied and the lowest unoccupied bands in the minority spins. 
The line compound is half-metallic, if $E_{-}\le 0$ and $E_{+}\ge 0$ (consequently, $n_{-}\le 0$ and $n_{+}\ge 0$). 
For half-metals with off-stoichiometric disorder,
the level of doping in terms of $n$ (added electrons per formula unit) lies within the range $[n_{-}, n_{+}]$.  
As explained in section~\ref{GuidanceRange}, dopants narrow the gap, hence the actual compositional range can be narrower. 
Importantly, our data allows to adjust composition towards $E_F$ in the middle of the gap.

%
\subsection{\label{Disorder}Substitutionally Disordered Alloys}
{\par} In materials engineering, considering multi\-component alloys with a partial  disorder can be challenging.  
Below we analyze quantitatively selected fH alloys, which have three sublattices (Fig.~\ref{fig1str}), and substitutional alloying is possible on each.  We illustrate this with solid-solution fH alloys with disorder on one sublattice and validate predictions, based on the electron count. 
We consider  Co$_{2}$Mn(Sn$_{x}$Sb$_{1-x}$), (Co$_{x}$Ni$_{1-x}$)$_{2}$MnA,  and Co$_{2}$(Mn$_{1-x}$Fe$_{x}$)A with A=\{Sn or Sb\}. 

\begin{figure}[t]
\centering{}\includegraphics[width=70mm]{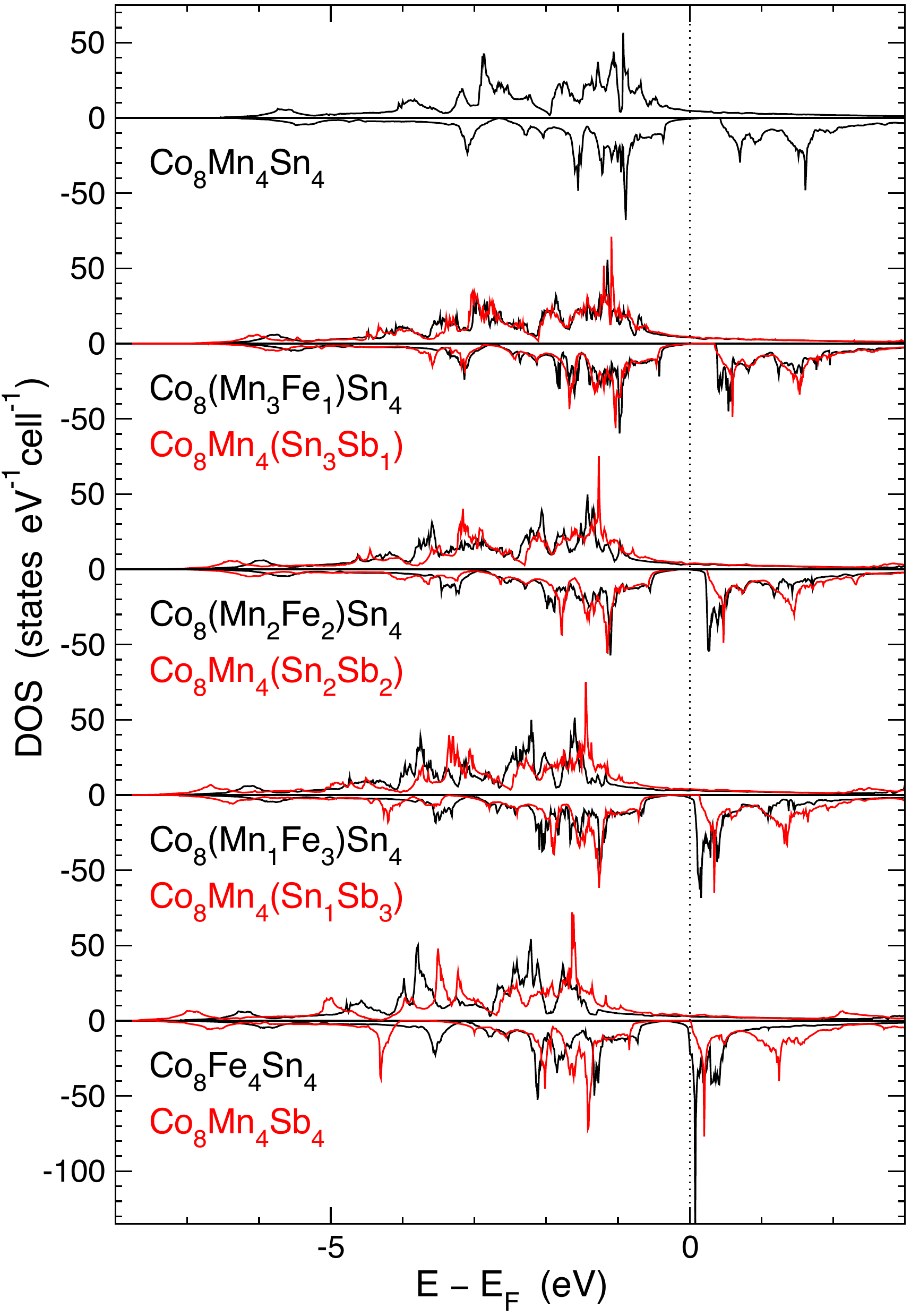} 
\caption{\label{Co2MnSnSb_Co2FeMnSn_DOS}  Total spin-resolved DOS of Co$_{8}$Mn$_{4}$(Sn$_{4x}$Sb$_{4-4x}$)
and Co$_{8}$(Mn$_{4-4x}$Fe$_{4x}$)Sn$_{4}$ fully-ordered Heusler with 16-atom cell. }
\end{figure}

\begin{figure}[t]
\centering{}\includegraphics[width=75mm]{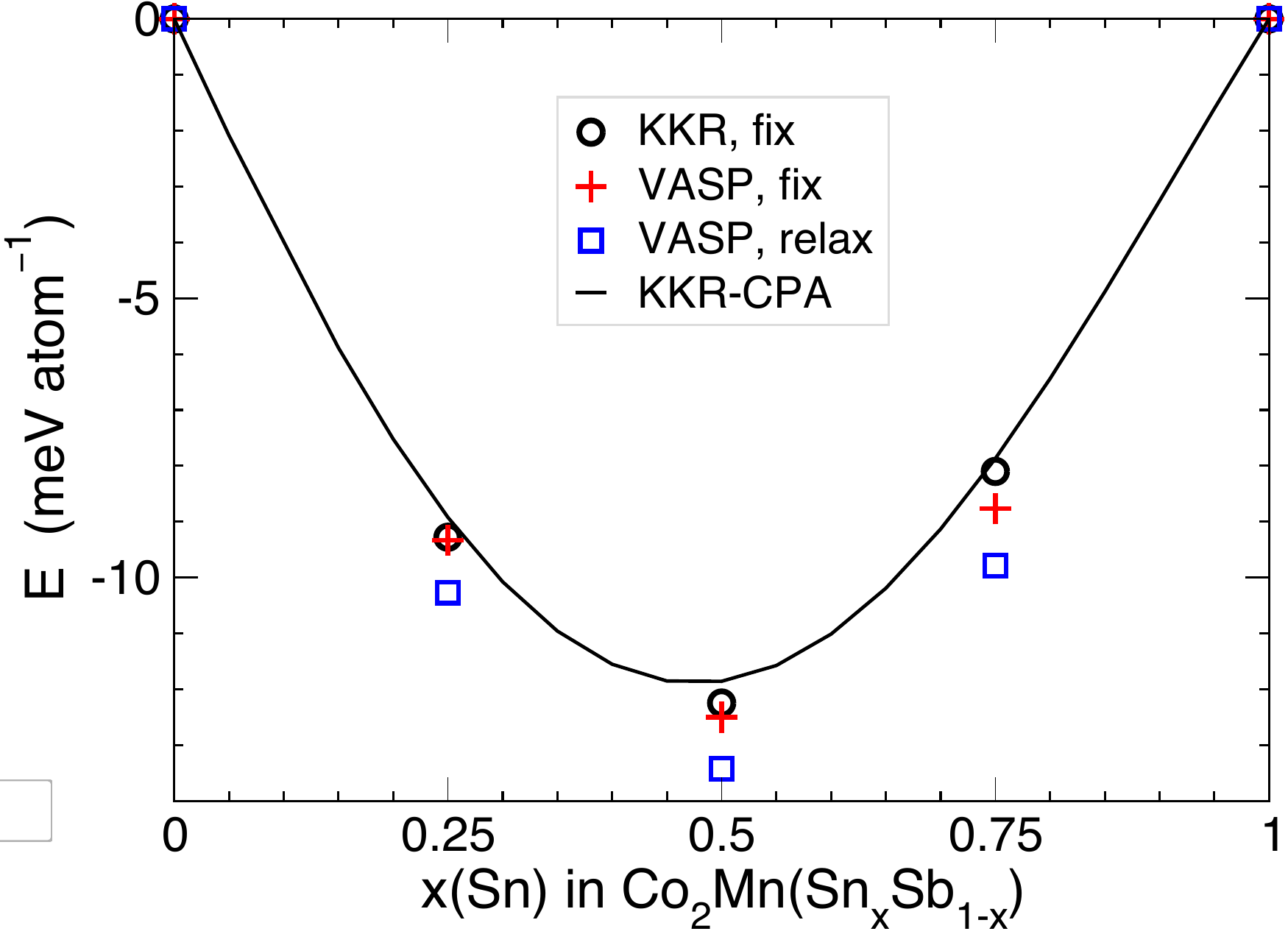}
\caption{\label{Co2MnSnSb_E} Co$_{2}$Mn$_{1}$(Sn$_{x}$Sb$_{1-x}$) mixing
enthalpies (meV/atom) at 0 K  and 0 Pa. Calculations with fixed atomic coordinates (fix) 
are compared to those with relaxed atoms (relax). Fully-ordered structures with 16-atom
cubic cells are addressed by VASP and KKR; KKR-CPA handles disorder on the (Sn,Sb) 
sublattice in the 4-atom cell shown in Fig.~\ref{fig1str}.}
\end{figure}

\begin{figure}[t]
\centering{}\includegraphics[width=75mm]{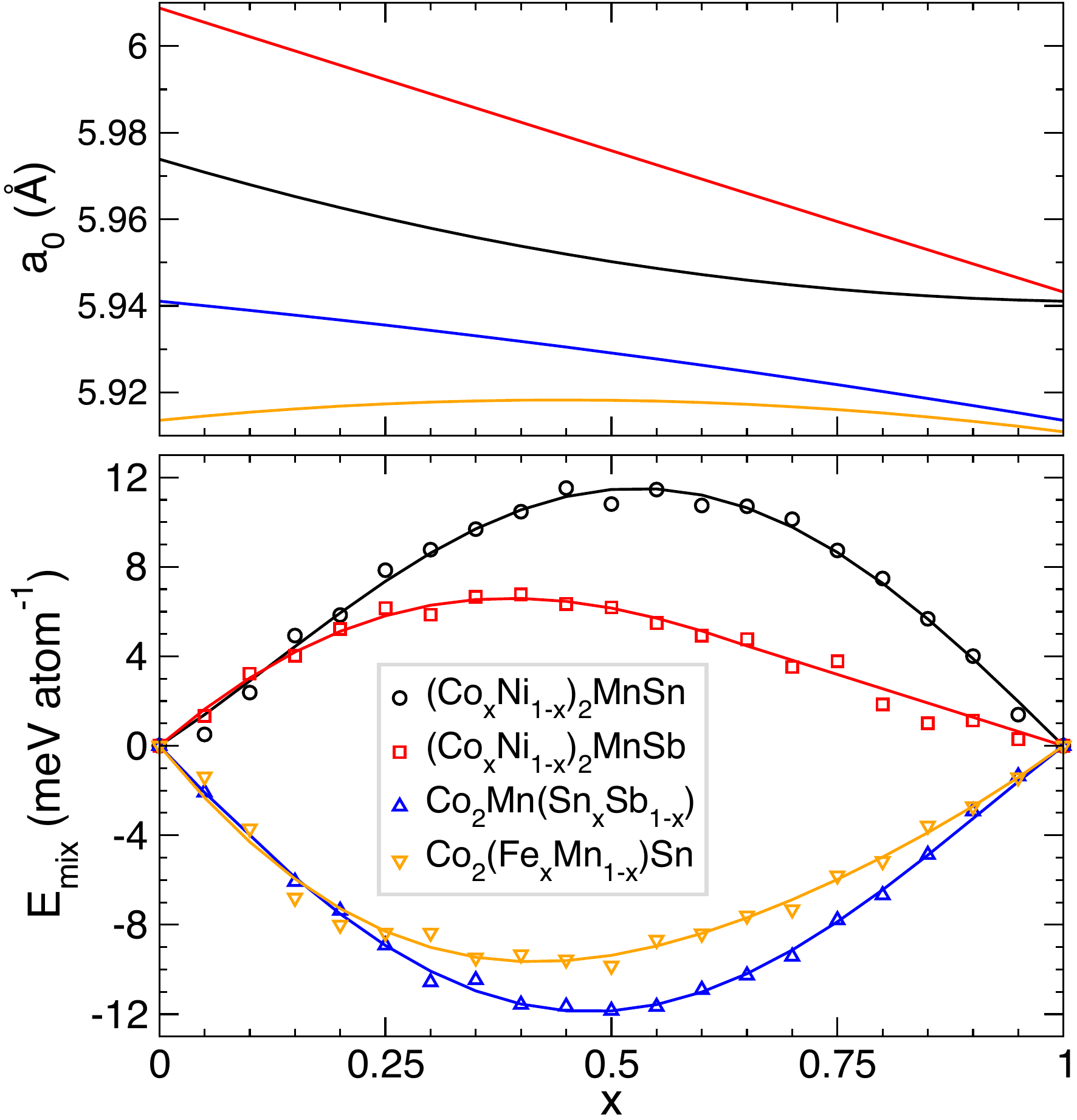} 
\caption{\label{Co2FeMnSn_E} The equilibrium lattice constants $a_{0}$~({\angstrom}) and mixing enthalpies $E_{mix}$ (meV/atom) vs. composition $x$ for fH alloys with a homogeneous chemical disorder on one of the sublattices obtained using KKR-CPA. 
Lines are the piece-wise low-degree polynomial fits to DFT data (21 points, symbols). 
} 
\end{figure}

{\par}  We emphasize that disorder on different sublattices can produce a similar shift of $E_{F}$.
Indeed, the expected similarity of the electronic structure of  Co$_{2}$Mn(Sn$_{x}$Sb$_{1-x}$) and Co$_{2}$(Mn$_{1-x}$Fe$_{x}$)Sn 
is verified in Fig.~\ref{Co2MnSnSb_Co2FeMnSn_DOS}. 
Notably, we find that their mixing enthalpies $E_{mix}$  are small compared to $k_{B}T_{0}$~=~$23.55$~meV ($k$ is Boltzmann's constant)
at temperature $T_0 \! = \! 273.15\,$K, see Figs.~\ref{Co2MnSnSb_E} and \ref{Co2FeMnSn_E}.

{\par} Solid-solution alloys with a small negative mixing enthalpy ($-k T_0 \! < \! E_{mix} \! \le \! 0$) 
can be uniform at ambient $T \ge T_0$.  In contrast, alloys with a small positive mixing enthalpy 
($0 \! < \! E_{mix} \! < \! kT_{0}$)  can develop a compositional fluctuation,  which lowers the 
enthalpy \cite{JPhysD51n2p024002y2018};  however, they do not segregate at $T > E_{mix} / k$.  
We find that  (Co$_{x}$Ni$_{1-x}$)$_{2}$MnA  with A=\{Sn or Sb\} 
have small positive $E_{mix}$,  see Fig.~\ref{Co2FeMnSn_E}. 

\begin{figure}[t]
\centering{}\includegraphics[width=75mm]{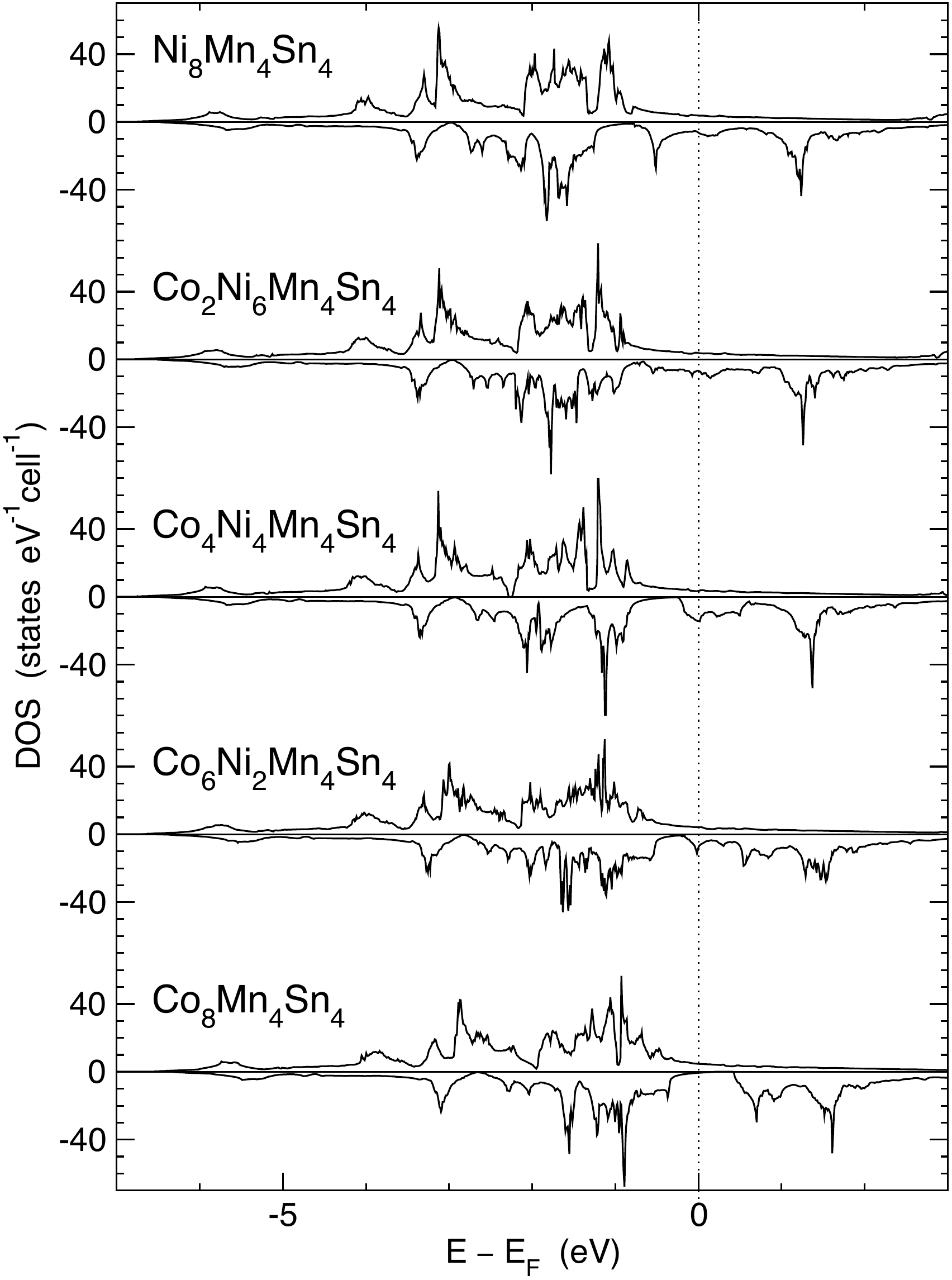} 
\caption{\label{CoNiMnSn_DOS} Total spin DOS (states/eV$\cdot$f.u.) of (Co$_{2-x}$Ni$_{x}$)MnSn.
cubic Heusler structures with 16-atom unit cell.
}
\end{figure}
\begin{figure}[t]
\begin{center}
\includegraphics[width=75mm]{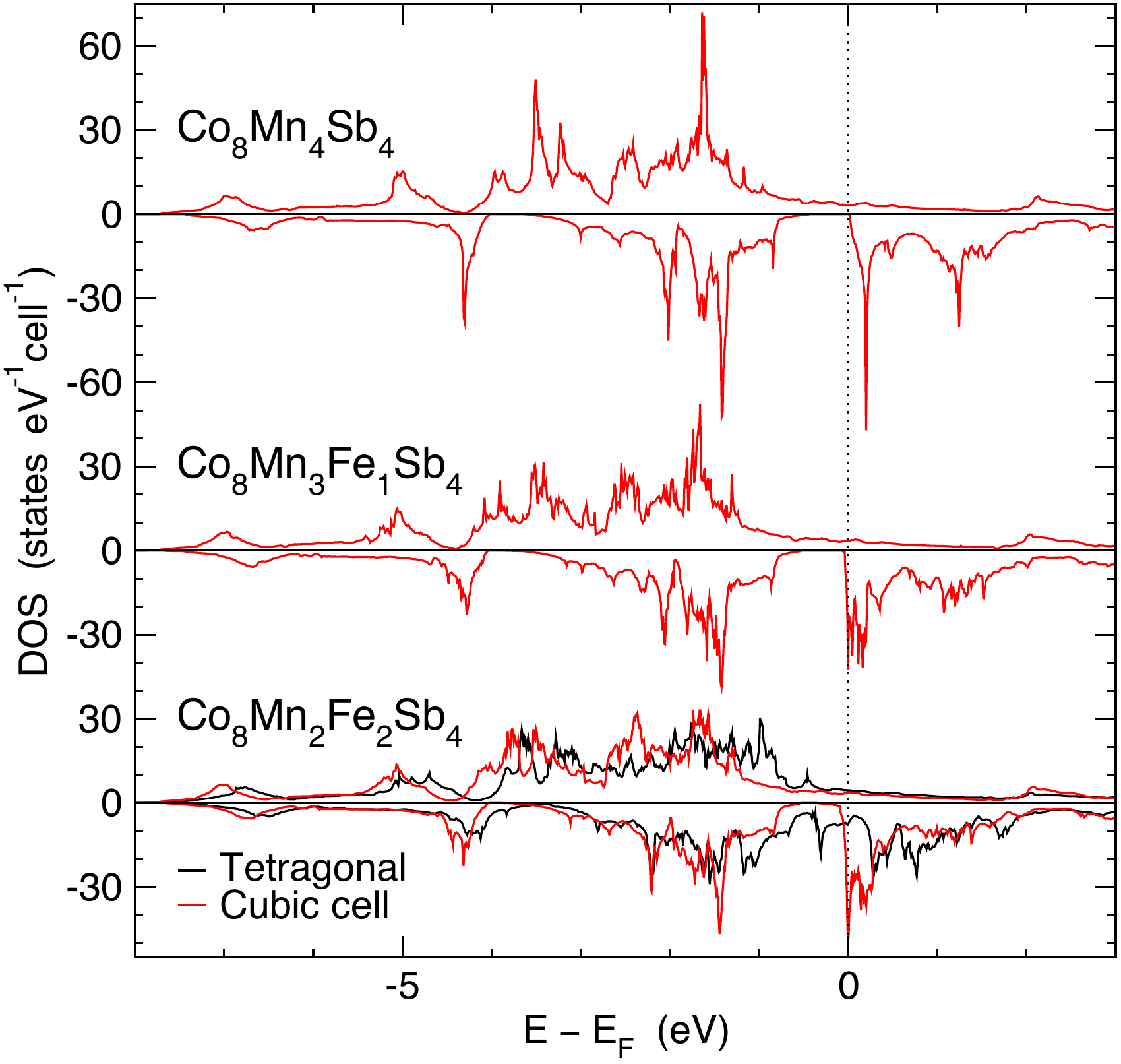}
\caption{\label{Co2FeMnSb_DOS3}   Total spin DOS (states/eV-f.u.) of Co$_8$(Mn$_{4-4x}$Fe$_{4x}$)Sb$_4$ 
at $0 \le x \le 0.5$ for cubic Heusler (red) and tetragonal (black) 16-atom unit cell.}
\end{center}
\end{figure}

\textbf{Co$_{2}$Mn(Sn$_{1-x}$Sb$_{x}$)}:  The calculated magnetization of Co$_{2}$Mn(Sn$_{1-x}$Sb$_{x}$) increases  with $x$ approximately linearly  from $1.258\,\mu_{B}$/atom in Co$_{2}$MnSn to $1.500\,\mu_{B}$/atom in half-metallic Co$_{2}$MnSb. The equilibrium lattice constants are slightly smaller than those from the Hume-Rothery rule: this indicates the mixing tendency. 

{\par} Indeed, the trend in mixing enthalpies, calculated   for the partially-ordered (KKR-CPA) and fully-ordered (VASP and KKR) systems, 
 is the same, see Fig.~\ref{Co2MnSnSb_E}.   The predicted stability of  disordered Co$_{2}$Mn(Sn$_{x}$Sb$_{1-x}$) alloys at room $T$ agrees with experiment \cite{JAP105n1p013716y2009}. The proposed compositional range of half-metals (given by $n$ in Table~\ref{t2n}) is confirmed by direct DFT calculations  in Fig.~\ref{Co2MnSnSb_Co2FeMnSn_DOS},  which shows that  disorder on different sublattices, resulting in the same change of $n$,
 can produce similar effects. 

{\par} Comparing calculations to experiment, we need to take into account that experimental samples are not always precisely stoichiometric. 
In particular,  according to the RBS measurements \cite{JAP105n1p013716y2009}, their Co$_{2}$MnSn sample had a Co-excessive composition Co$_{2.03}$Mn$_{1.00}$Sn$_{0.97}$, while Co$_{2}$MnSb sample was Co-deficient Co$_{1.98}$Mn$_{1.01}$Sb$_{1.01}$
(Co$_{1.94}$Mn$_{1.02}$Sb$_{1.04}$ according to ICP).  
Consequently, we expect a slight difference between the equilibrium lattice constants $a$, calculated for the ideal stoichiometric crystals at $0\,$K, 
and measured on the off-stoichiometric samples at finite $T$.  The calculated and experimental \cite{Buschow1983,JAP105n1p013716y2009} 
lattice constants 
agree within 1\%, see Table~\ref{t2c}.

\textbf{(Co$_{2-x}$Ni$_{x}$)MnSb}:  With increasing \%Ni, magnetization decreases and $a$ is increasing (Fig.~\ref{Co2FeMnSn_E}). As  replacement of Co by Ni in Co$_{2}$MnSb moves $E_{F}$ away from the gap, (Co$_{2-x}$Ni$_{x}$)MnSb are not half-metals at $x>0.03$,  see Table~\ref{t2n}.

\textbf{(Co$_{2-x}$Ni$_{x}$)MnSn}: These alloys can be half-metallic at $0.1 \! < \! x \! < \! 0.45$, see Table~\ref{t2n}.    
The rapidly quenched by melt-spinning Ni$_{47.3}$Mn$_{28.2}$Sn$_{24.5}$ sample was confirmed to be half-metallic \cite{JMMM386p98p2015}:  
this sample has excessive 2.7 at.\%Mn replacing Ni and 0.5 at.\%Mn replacing Sn. 

{\par} Figure~\ref{CoNiMnSn_DOS}  shows that doping not only shifts  $E_{F}$, but also narrows the gap.  Due to the peak of DOS at the $E_{F}$ (Fig.~\ref{CoNiMnSn_DOS}), cubic CoNiMnSn and Co$_3$NiMn$_2$Sn$_2$ structures 
are unstable. We find that they can be stabilized by a tetragonal distortion, and our findings agree with recent calculations \cite{JPhysD49n39p395001y2016}. 

\textbf{Co$_{2}$(Mn$_{1-x}$Fe$_{x}$)Sn}: In Co$_{2}$MnSn, the gap in minority spins is located slightly above $E_{F}$. 
Shifting $E_{F}$ to higher energies should change this alloy into a half-metal, see Fig.~\ref{Co2MnSnSb_Co2FeMnSn_DOS}. 
Magnetization increases with $x$ from $1.26\,\mu_{B}$/atom in Co$_{2}$MnSn to $1.41\,\mu_{B}$/atom in Co$_{2}$FeSn and so does the 
$a$ (although the atomic radius and magnetic moment of Mn is larger than that of Fe).

\textbf{Co$_{2}$(Mn$_{1-x}$Fe$_{x}$)Sb}: In Co$_{2}$MnSb the $E_{F}$ is located at the gap's edge in minority spins (Fig.~\ref{fig5dos}). 
And, Co$_2$(Mn$_{1-x}$Fe$_x$)Sb are not half-metals for $x>0.0$. 
Replacement of Mn by Fe lowers magnetization from $1.50\,\mu_{B}$ in Co$_{2}$MnSb to $1.32\,\mu_{B}$/atom in Co$_{2}$FeSb, while distortion of Co$_{2}$(Mn$_{0.5}$Fe$_{0.5}$)Sb from unstable cubic to a more stable tetragonal structure lowers magnetization from 1.42 to $1.23\,\mu_{B}$/atom. 
An attempt to mix Mn with Fe on the Mn sub\-lattice quickly moves $E_{F}$ away from the band gap  into the large peak in the minority spin DOS (Fig.~\ref{Co2FeMnSb_DOS3}), making a cubic Heusler structure unstable. 
Indeed, mixing enthalpies of cubic Co$_{2}$(Mn$_{1-x}$Fe$_{x}$)Sb  are positive (relative to the segregated  Co$_{2}$MnSb and Co$_{2}$FeSb), with that for cubic Co$_{2}$(Mn$_{0.5}$Fe$_{0.5}$)Sb being $+2.1\,$meV/atom, while distorted tetragonal Co$_{2}$(Mn$_{0.5}$Fe$_{0.5}$)Sb has a negative formation enthalpy of $-10.3\,$meV/atom.

A similar instability towards tetragonal distortions was predicted for other  Co$_2$-based Heusler compounds under pressure \cite{JPhysD49n35p355004y2016}. We conclude that adding iron to Co$_{2}$MnSb can result in formation of other (more stable) compounds with lower magnetization, which are not Heusler.


\subsection{Further Comparison to Experiment}
\begin{table}[b]
\caption{\label{t2c} Equilibrium lattice constant $a$ ({\angstrom}),
magnetic moment $M$ ($\mu_{B}$/f.u.), and experimental \cite{Buschow1983,PJW1971} Curie temperature
$T_{c}$ (K) of fully relaxed fH Co$_{2}$MnX alloys from theory
(DFT: GGA, see Appendix~\ref{subsec:VASP}) and experiment (Expt.). Asterisk ($^{*}$) marks hypothetical
non-existent compounds. 
Known competing structures include oP12 \cite{TiNiSi1998} for X=\{Si, As\} and tI12 \cite{Matveyeva1968} for Co$_{2}$MnSn.
}
\begin{tabular}{|l|c|c|c|l|ll|}
\hline 
\multirow{2}{*}{X } & \multicolumn{2}{c|}{\emph{a(}{\angstrom}\emph{) }} & \multicolumn{2}{c|}{ \emph{M(}$\mu_{B}$/\emph{f.u.)} } & \emph{$T_{c}$(K)}  &  \tabularnewline
\cline{2-7} 
 & DFT  & Expt. \cite{Buschow1983,PJW1971}  & DFT  & Expt.   & Expt.   &  \tabularnewline
\hline 
Al  & 5.6927  & 5.749\cite{Buschow1983}, 5.756\cite{PJW1971}  & 4.03  & 4.04, 4.01 & 693 & \tabularnewline
Ga  & 5.7136  & 5.767\cite{Buschow1983}, 5.770\cite{PJW1971}  & 4.09  & 4.05, 4.05 & 694  & \tabularnewline
In$^{*}$  & 5.9813  & \textendash{}  & 4.44  & \textendash{} & \textendash{} & \tabularnewline
Tl$^{*}$  & 6.0561  & \textendash{}  & 4.80  & \textendash{} & \textendash{} & \tabularnewline
\hline 
Si  & 5.6285  & 5.645\cite{Buschow1983}, 5.654\cite{PJW1971} & 5.00  & 4.90, 5.07 & 985 &  \tabularnewline
Ge  & 5.7358  & 5.749\cite{Buschow1983}, 5.743\cite{PJW1971} & 5.00  & 4.93, 5.11& 905  & \tabularnewline
Sn  & 5.9854  & 5.984\cite{Buschow1983}, 6.000\cite{PJW1971} & 5.03  & 4.79, 5.08 & 829  &  \tabularnewline
Pb$^{*}$  & 6.0956  & \textendash{}  & 5.11  & \textendash{} & \textendash{} & \tabularnewline
\hline 
P$^{*}$  & 5.6385  & \textendash{}   & 6.00  & \textendash{} & \textendash{} & \tabularnewline   
As$^{*}$  & 5.7939  & \textendash{}  & 5.99  & \textendash{} & \textendash{} &  \tabularnewline
Sb  & 6.0182  & 5.943\cite{JAP105n1p013716y2009}  & 6.00 & 4.52\cite{Buschow1983}  & \textendash{} & \tabularnewline
Bi$^{*}$  & 6.1793  & \textendash{}  & 6.00  & \textendash{} & \textendash{} & \tabularnewline
\hline 
\end{tabular}
\end{table}

{\par} Addressing the ternary line compounds, we fully relaxed each structure (Appendix~\ref{subsec:VASP}). We found a reasonable agreement
between the calculated and measured \cite{Buschow1983,PJW1971} lattice constants $a$ and moments $M$, see Tables~\ref{t2c} and \ref{t3a}. 
 We considered both ferromagnetic (FM) and antiferromagnetic (AFM)  spin ordering. For the Co$_{2}$MnX systems in Fig.~\ref{fig5dos},
we find that FM ordering is preferred, in agreement with the previous  calculations \cite{PRB28p1745y1983}. 

{\par}The calculated \cite{VASP1,VASP2} magnetization $M$ of Co$_{2}$MnSn is $5.03\,(5.04)\,\mu_{B}$
per formula unit  with Sn $d$-electrons in the core (valence) at the equilibrium $a_{0}=5.9854\,${\angstrom}. 
It reasonably agrees with the experimental $M$ of $5.08\,\mu_{B}$. \cite{Webster1969}  
Our  all-electron KKR results yield $4.98\,\mu_{B}$/f.u. at $a_{0}=5.995\,${\angstrom}, comparing well to earlier FLAPW \cite{FLAPW2002_Picozzi} of  $5.0\,\mu_{B}$/f.u. at $a_{0}=5.964\,${{\angstrom}.
However, we note that there is an experimental uncertainty in the weight of the powder sample, only a fraction of which is the desired phase \cite{Buschow1983}.  The imprecisely measured magnetization of the Co$_{2}$MnSn and Co$_{2}$MnSb  polycrystalline powder  samples \cite{Buschow1983} is  below the expected theoretical value, which must be integer for a half-metal,  and Co$_{2}$MnSb is shown to be a half-metal in experiment \cite{Buschow1983}.

\subsection{\label{hH}  hH-NiMnX alloys }
{\par}  Among the hH-NiMnX alloys, where X is one of \{Al, Ga, In, Tl; Si, Ge, Sn, Pb; P, As, Sb, Bi\}, NiMnSb is the only known stable ternary compound, which had been claimed to be a half-metal \cite{Groot1983,NatPhys12p855y2016}, but its half-metallicity was questioned by some measurements \cite{PRB68p104430y2003} and calculations \cite{PRB81p054422y2010}.

We calculate magnetization (Fig.~\ref{fig2M}), formation energy (Fig.~\ref{fig3E}), and electronic DOS (Figs.~\ref{fig5dos} and \ref{DOS_Ni2MnX_fH}) of hH--NiMnX and 
fH--Ni$_{2}$MnX.  Magnetization of the half-metallic hH alloys increases from 3 $\mu_{B}$/f.u. for NiMnSi and NiMnGe containing group 14 elements to 4 $\mu_{B}$/f.u. in group 15. An antiferromagnetic (AFM) spin ordering is preferred for most compounds of sulfur, a group 16 element with a small atomic size and high electronegativity.

\begin{figure}[t]
\includegraphics[width=75mm]{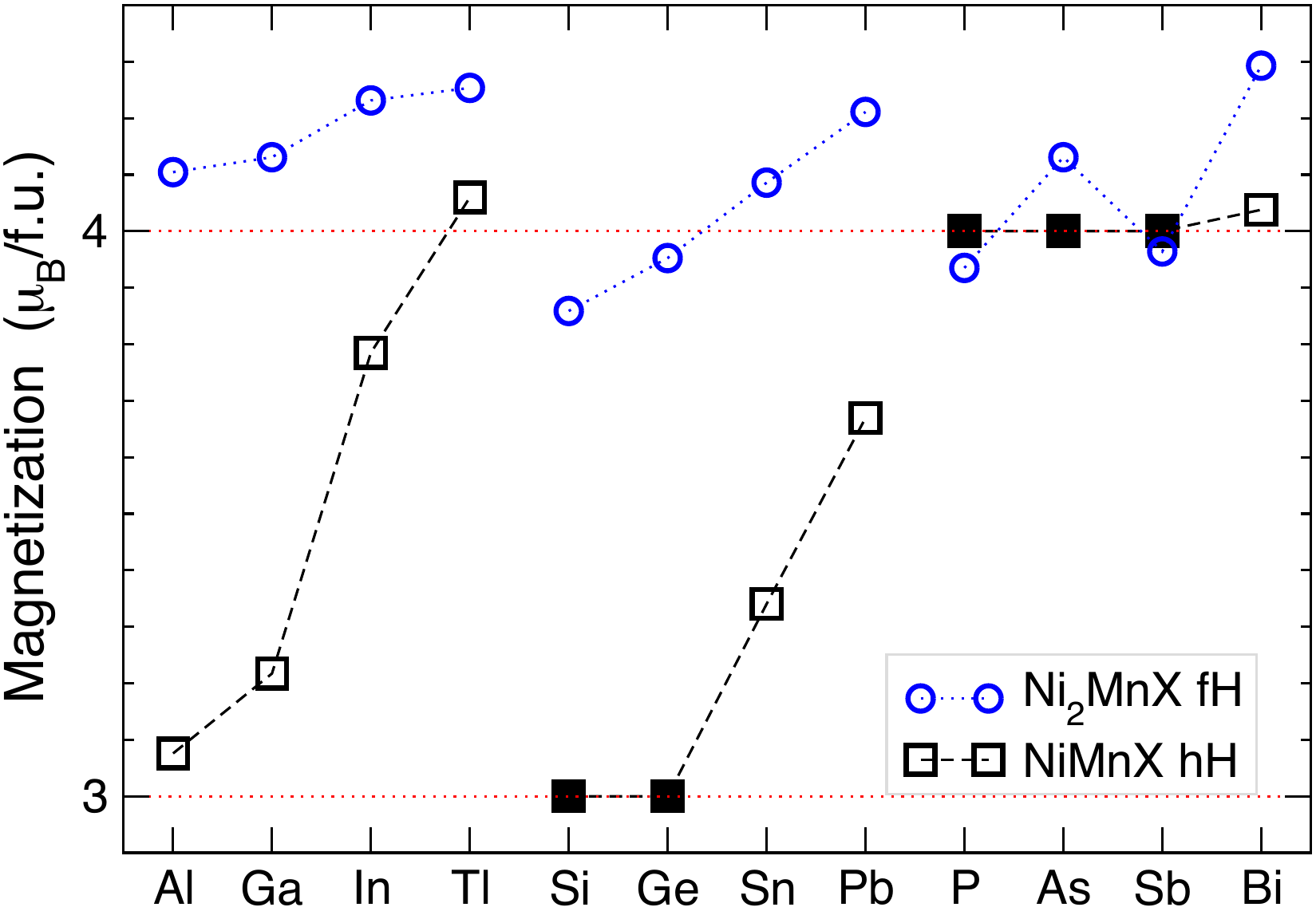} 
\caption{\label{fig2M} Calculated magnetization (Bohr magnetons per f.u.) for hH-NiMnX (squares); solid symbols mark  
 half-metals.  fH-Ni$_{2}$MnX (circles) are ferrimagnetic (see appendix \ref{Ni2MnX}).
}
\end{figure}
\begin{figure}[t]
\includegraphics[width=75mm]{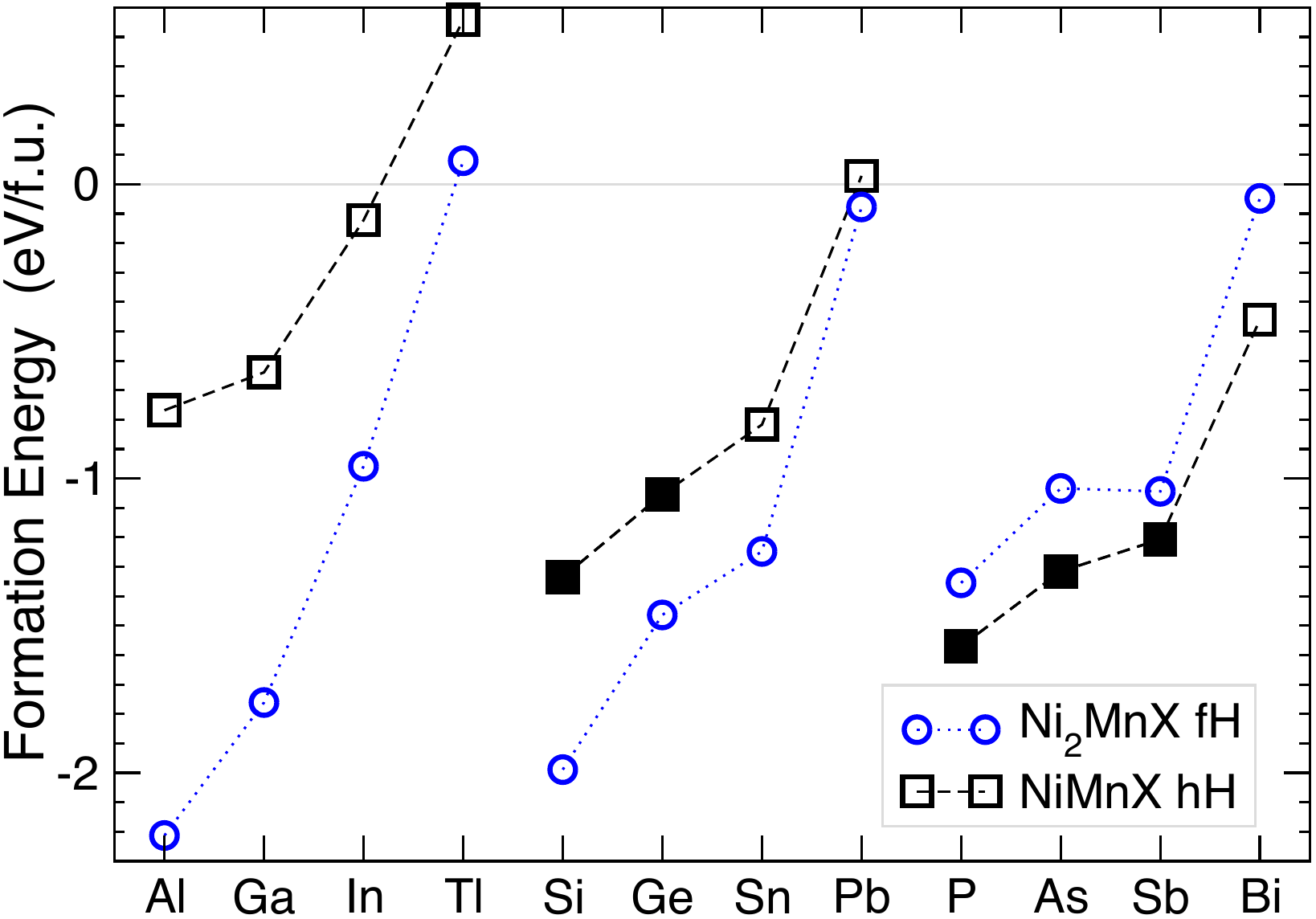} 
\caption{\label{fig3E} Formation energy (eV per f.u.) of hH-NiMnX (squares);  solid symbols mark  
 half-metals.
Results for fH-Ni$_{2}$MnX  (circles) are given for comparison (see appendix \ref{Ni2MnX}).}
\end{figure}

\begin{figure}[b]
\includegraphics[width=75mm]{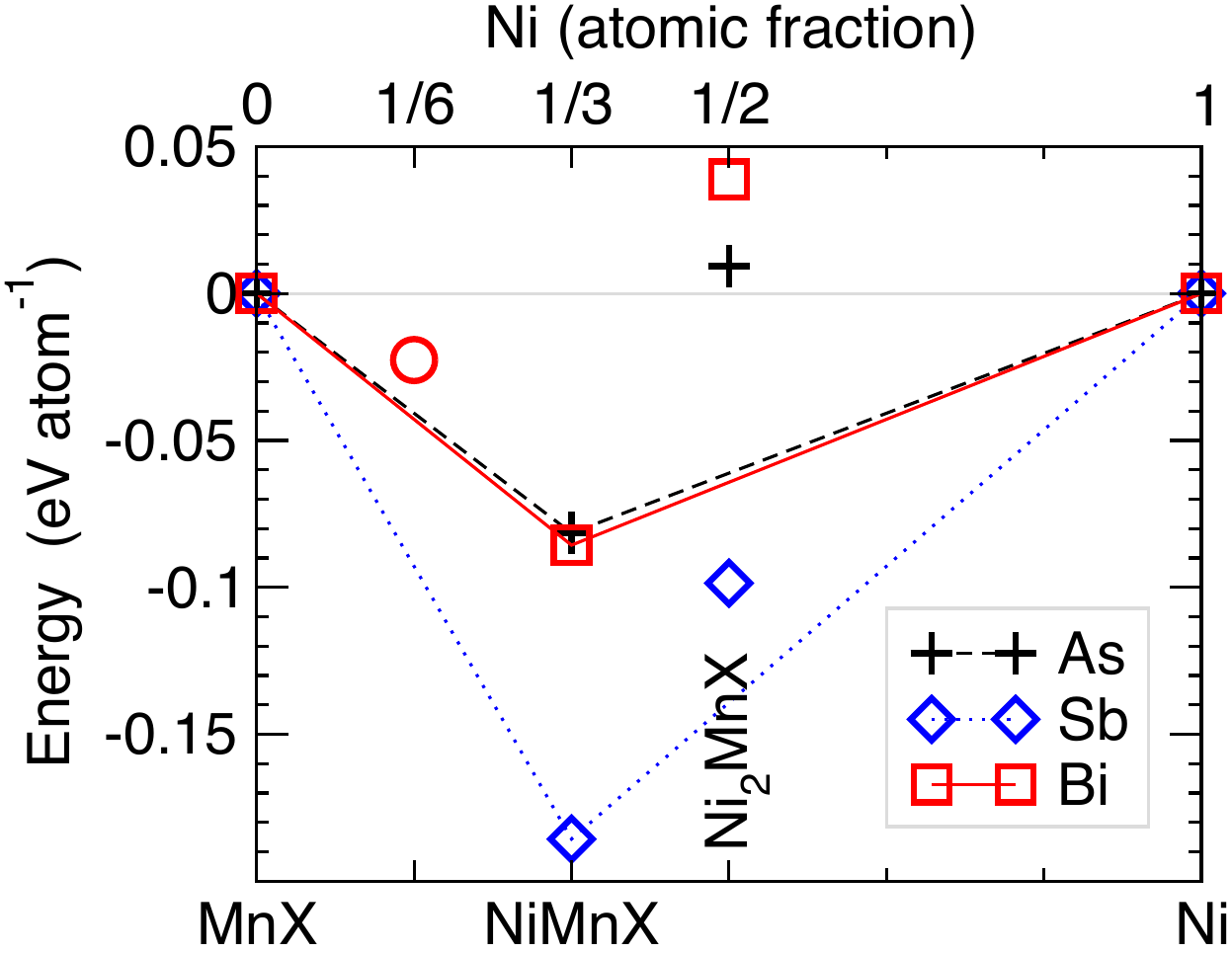} 
\caption{\label{fig4E2} 
Energy (eV/atom) of  hH-NiMnX relative to fcc Ni and MnX (X = As, Sb, Bi), see Fig.~\ref{fig_el}, and
 phase-segregated Ni$_{2}$Mn$_{5}$Bi$_{4}$ and metallic Bi (circle). Ground-state hull is denoted by lines. }
\end{figure}

\begin{table}[b]
\caption{\label{t3a} Lattice constant $a$ ({\angstrom}) of fully relaxed
 hH NiMnX alloys from theory (VASP--GGA) and
experiment \cite{Buschow1983,JMMM65p76y1987,Uhl1982}. 
Other observed competing structures are listed in the last column.
}
\begin{tabular}{|c|cc|c|}
\hline 
 \multirow{2}{*}{X } & NiMnX & hH  & Other \tabularnewline
 & GGA  & Expt.  & Expt. \tabularnewline

\hline 
Al  & 5.6078  & -  & \tabularnewline
Ga  & 5.6182  &  & \tabularnewline
In  & 5.9521  &  & \tabularnewline
Tl  & 6.0449  &  & \tabularnewline
\hline 
Si  & 5.44837  &  & oP12 \cite{TiNiSi1998} \tabularnewline
Ge & 5.570348  &  & oP12, hP6 \tabularnewline
Sn  & 5.8903  & -  & \tabularnewline
Pb  & 6.0412  &  & \tabularnewline
\hline 
P  & 5.46467  &  & oP12, hP9 \tabularnewline
As  & 5.637066  &  & oP12, hP9 \tabularnewline
Sb  & 5.9068  & 5.920 \cite{Buschow1983}  & NiMnSb$_{2}$ hP4 \tabularnewline
Bi  & 6.08546  &  & {[}cF88{]} \tabularnewline
\hline 
\end{tabular}
\end{table}

\begin{table}[hb]
\caption{\label{t1E} 
Formation energies (eV/atom) of weakly stable bismuth compounds
from theory and experiment.}
\begin{tabular}{|l|c|c|}
\hline 
 & E (Theory)  & E (Expt.) \tabularnewline
 & eV/atom  & eV/atom \tabularnewline
\hline 
MnBi  & -0.102  & -0.102 \tabularnewline
NiBi  & -0.064  & -0.020 \tabularnewline
Ni$_{2}$Mn$_{5}$Bi$_{4}$  & -0.117  & stable \tabularnewline
CuNi$_{3}$Mn$_{4}$Bi$_{4}$  & -0.106  & --- \tabularnewline
ZnNi$_{7}$Mn$_{8}$Bi$_{8}$  & -0.115  & --- \tabularnewline
\hline 
\end{tabular}
\end{table}

\begin{figure}[t]
\includegraphics[width=80mm]{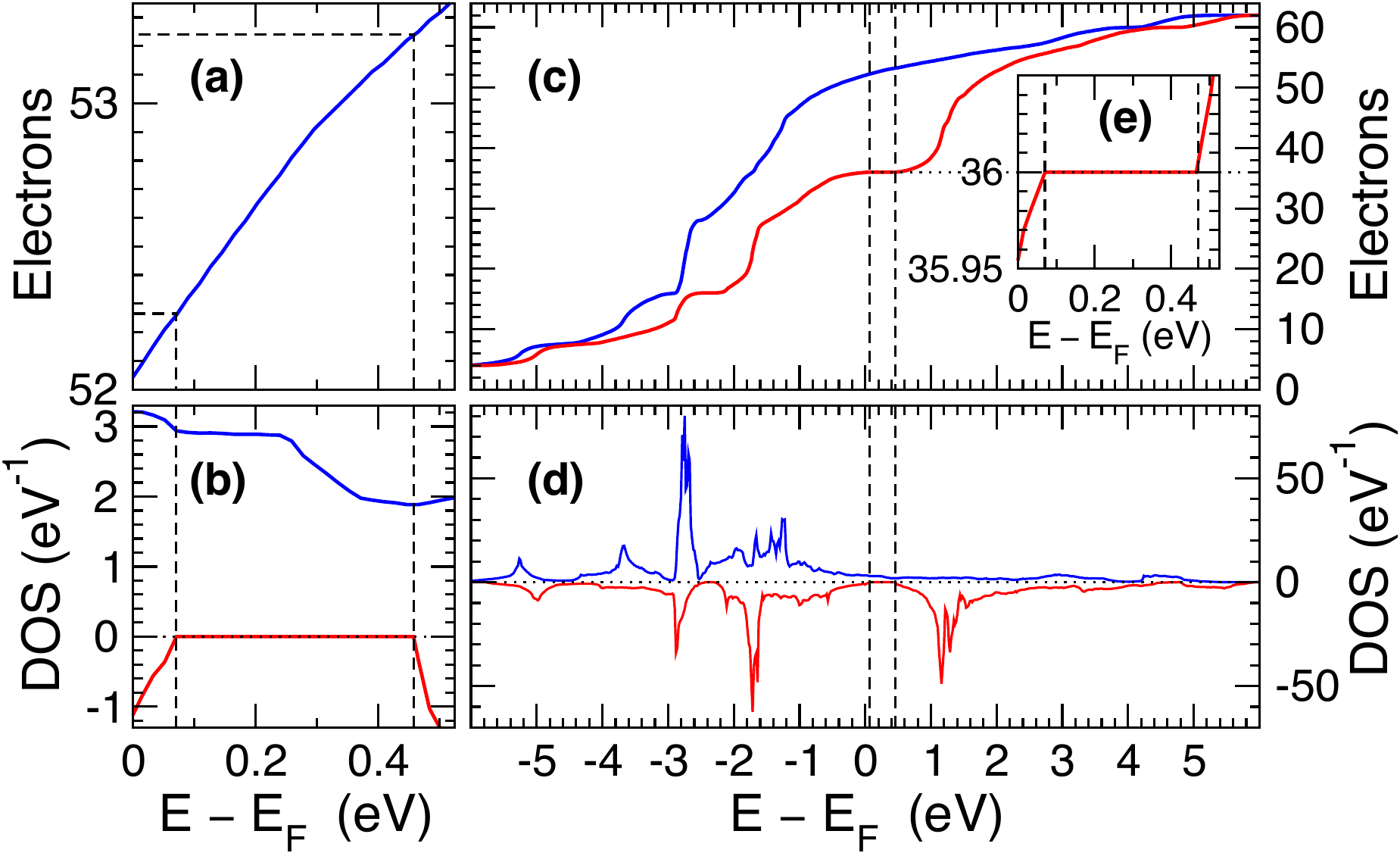} \caption{\label{fig6el} 
Number of electrons (per unit cell) and DOS (states/eV$\cdot\,$cell) of the hH-NiMnBi. }
\end{figure}

Among hH alloys, we find five half-metals: they are NiMnX with X=\{Si, Ge, P, As, Sb\}. The first two with a  group 14 element (NiMnSi, NiMnGe) have magnetization of 3 $\mu_{B}$/f.u., which increases to 4 $\mu_{B}$/f.u. for the last three (NiMnP, NiMnAs, NiMnSb) with X from group 15.
In spite of having large magnetic moments, cubic systems cannot make hard magnets due to absence of anisotropy;  a structural anisotropy is necessary for a magnetic anisotropy in a hard magnet.

{\par} All considered hH--NiMnX alloys have a gap in the minority spins at or slightly above $E_F$ (Fig.~\ref{fig5dos} and Table~\ref{t2n}). In general, a peak with a maximum in the DOS at  $E_{F}$ destabilizes the alloy.  Electron (or hole) doping is one option, which shifts $E_{F}$ away from this peak, and reduces $g(E_{F})$.  Half-metals are the most stable compounds among the considered hH alloys (Fig.~\ref{fig3E}).

{\par}Figure~\ref{fig4E2} shows relative energies of the line compounds.   
The fH--Ni$_2$MnX alloys with X=\{As, Bi\} are not stable, because their energies are well above the ground-state hull.  
Although the fH Ni$_2$MnSb could be stabilized by entropy at finite temperature $T$, we predict that at low $T$ it 
tends to segregate towards Ni$_{2 - \delta }$MnSb and metallic Ni, although diffusion is limited at low $T$. 
The hH--NiMnSb is known to be stable in experiment, and Fig.~\ref{fig4E2} does not question its stability. 

{\par} MnNiBi is only 0.06 $e^{-}$/f.u. away from being a half-metal;  it  has a gap in the minority-spin DOS above  $E_{F}$ (Fig.~\ref{fig6el}), 
and might be turned into a half-metal by electron doping, such as a partial substitution of Ni by Cu or Zn (Fig.~\ref{fig7dos4}).
From the estimate of the needed level of doping (see Fig.~\ref{fig6el}  and Table~\ref{t2n}), we predict that CuNi$_{3}$Mn$_{4}$Bi$_{4}$ and ZnNi$_{7}$Mn$_{8}$Bi$_{8}$ should be half-metals.   Our supercell calculations (Fig.~\ref{fig7dos4})  confirm this prediction, although doping narrows the band gap in the minority spins,  as discussed in section~\ref{Guidance}.


\begin{table}[b]
\caption{\label{t2n2} For selected compounds, $[n_{-},n_{+}]$ range ($e^{-}$/f.u.),
$E_{-}$, $E_{+}$, and $E_{gap}=E_{+}-E_{-}$ (eV) from the supercell
calculations, see Table~\ref{t2n} caption.
}
\begin{tabular}{rllllll}
\hline 
 & $n_{-}$  & $n_{+}$  & $E_{-}$  & $E_{+}$  & $E_{gap}^{GGA}$  & $E_{gap}^{vLB}$ \tabularnewline
\hline 
NiMnBi                                      & 0.06  & 0.31  & 0.07 & 0.47  & 0.40  & 0.34 \tabularnewline
CuNi$_{3}$Mn$_{4}$Bi$_{4}$  &-0.57  & 1.1 & -0.10 & 0.25 & 0.35 & 0.25 \tabularnewline
ZnNi$_{7}$Mn$_{8}$Bi$_{8}$  & -0.12 & 0.2  & -0.11 & 0.19 & 0.30 &--- \tabularnewline
\hline 
\end{tabular}
\end{table}

\begin{figure}[t]
\includegraphics[width=75mm]{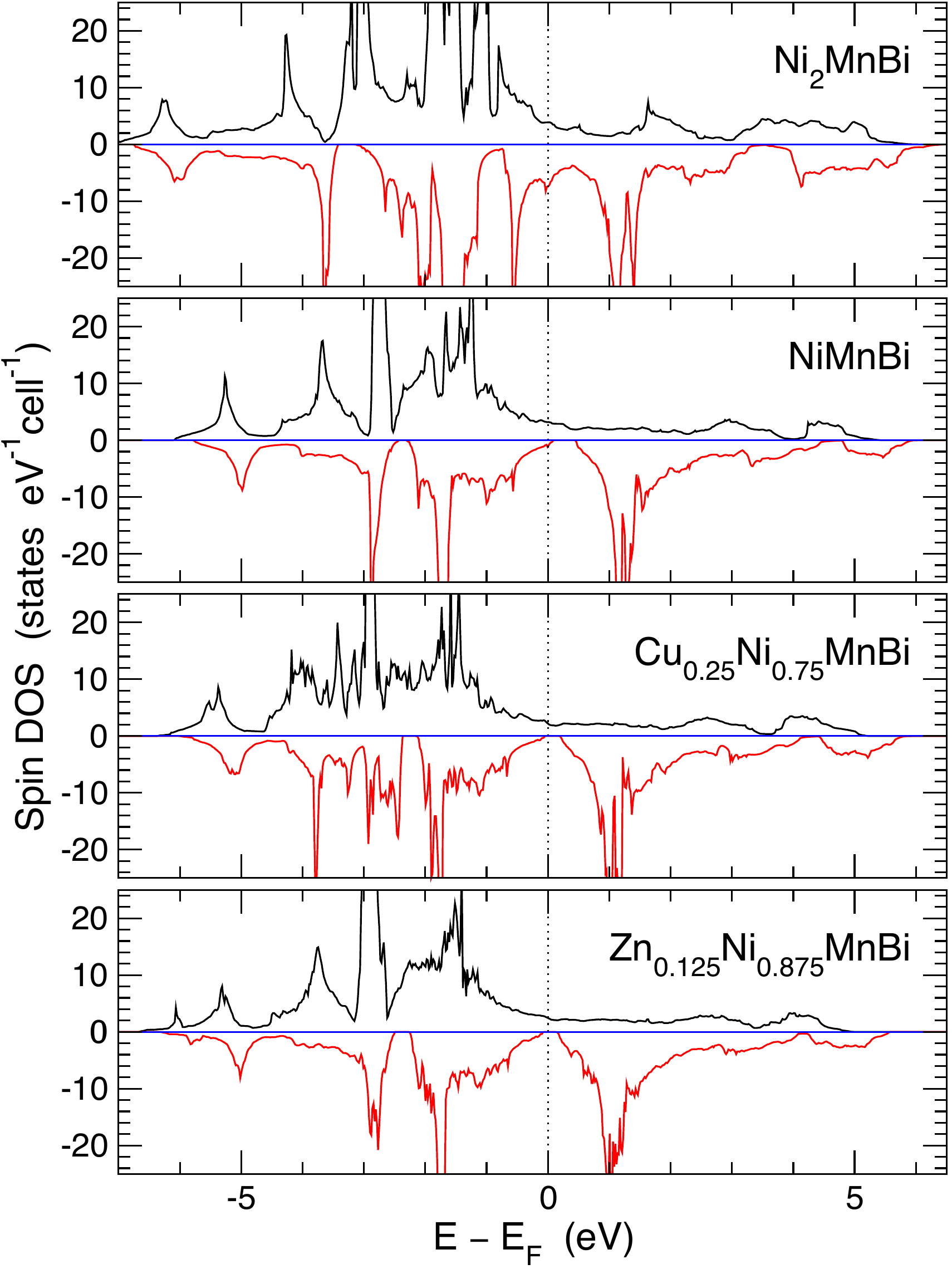}
 \caption{\label{fig7dos4} Total spin DOS (states/eV-f.u.) of fH-Ni$_{2}$MnBi (unstable), hH-NiMnBi, and 
 doped hH alloys with substituted N, which shifts $E_{F}$ and narrows the gap in minority spins. 
}
\end{figure}

{\par} 
Table~\ref{t4n} can be used to design half-metals of composition
(Ni$_{1-z-y}$Cu$_{z}$Zn$_{y}$)MnX with X=\{Al, Ga, In, Tl; Si,
Ge, Sn, Pb; P, As, Sb, Bi\}. For each X, we predicted a range of $n=(z+2y)$,
at which this doped half-Heusler alloy is half-metallic. For negative
$n$, Ni can be mixed with Co and Fe to make (Ni$_{1-z-y}$Co$_{z}$Fe$_{y}$)MnX
alloys, where $(z+2y)=|n|>0$. Dopants narrow the band gap, hence
the actual composition range could be narrower than that predicted
from electron count.

\subsection{\label{Stability}Stability}

{\par }  In general, half-metals are more stable than other metallic half-Heusler alloys, see Fig.~\ref{fig3E}.
 We emphasize that a negative formation energy (Fig.~\ref{fig3E}) relative to the elemental ground states does not necessarily indicate stability of a given phase, because there could be other (more stable) phases nearby, such as the weakly stable Ni$_{2}$Mn$_{5}$Bi$_{4}$, NiBi and MnBi in Ni-Mn-Bi system (Fig.~\ref{fig_el} and Table~\ref{t1E}).
Energies of all possible competing structures are needed for constructing the complete ground-state hull. While an alloy with a negative formation
energy might be stable, a positive energy relative to any set of phases indicates that a structure is above the ground-state hull and is definitely unstable towards segregation at low $T$. However, some of the metastable structures are technologically feasible, especially if the phase transformation or segregation has high enthalpy barriers or a metastable phase is thermally stabilized by entropy; examples are iron steels \cite{Bugayev1971,PRB91p174104,JChemPhys142p064707y2015},
Alnico magnets \cite{Campbell1994}, titanium alloys \cite{PRB93p020104y2016},
shape memory alloys \cite{PRL113p265701y2014,PRB90p060102y2014,C2NEB},
and graphite \cite{JChemPhys38n3p631y1963}.

\subsection{\label{Discussion}Predicted Weakly Stable Bismuthides}
The ferrimagnetic hexagonal MnBi, NiBi, and magnetic cubic Ni$_{2}$Mn$_{5}$Bi$_{4}$ are weakly stable compounds. 
In particular,  determined from the measured heat of combustion \cite{Shchukarev1961} formation  enthalpy of MnBi is only  $-4.7\pm0.1\,$kcal/g-formula ($-0.102\pm0.002\,$eV/atom). 
The measured formation enthalpy of NiBi is approximately $-0.93\,$kcal/g-formula  ($-0.020\,$eV/atom); this value was inaccurate due to inhomogeneity of the sample \cite{Predel1972}.  The calculated formation enthalpies of NiBi and Ni$_{2}$Mn$_{5}$Bi$_{4}$ reasonably agree with experiment (Table~\ref{t1E}).  Interestingly, we predict a new family of weakly stable half-metallic bismuthides.  We look forward towards production  of the doped multi\-component  half-metals, constructed using  electron count in Tables~\ref{t2n},  \ref{t4n} and \ref{t2n2}.

\section{\label{Summary}Summary}
{\par} If a material is a half-metal, then there exists a range of half-metallic compositions, which can be expressed in terms of the electron count. 
Within that compositional range, the Fermi energy remains within the band gap in one of the spin manifolds. 
To offer a guidance for the band-gap engineering, we predict half-metallic compositional ranges for the considered full- and half-Heusler alloys. 
The key take away is that this is a general methodology to tune materials to a nearby gap.

{\par} We addressed both electronic properties and structural  stability in  a wide class of full- and half-Heusler compounds. 
 We found that many of the Heusler Co$_{2}$MnX alloys (with X=\{Al, Si, Ga, Ge, Sn, Sb\}) are stable,  have low mixing enthalpies  for disorder on each of 3 sublattices, and have a band gap in the minority spins at or near the Fermi energy. 
Within the predicted compositional ranges, these materials are half-metals.  
Considering effect of alloying on half-metallicity, 
we selectively verify electronic properties  by direct  calculations in supercells. 
Our results (see Tables~\ref{t2n} and \ref{t4n} and Fig.~\ref{fig5dos}) facilitate design of multicomponent half-metals (including those with off-stoichiometric chemical disorder)   with desired width and position of the spin band gap.

Compared to the searches for half-metals among the stoichiometric line compounds 
\cite{JMMM441p333y2017,CompMatSci148p87y2018,JMMM491p165560y2019,JMateriomics5n3p404y2019,JMMM491p165567y2019,PhysicaScripta94n12p125001y2019,PhysRevMaterials3p024410y2019,PhysRevMaterials3p124415y2019} (which constitute a finite countable set of discrete points in the compositional space),  
consideration of half-metallic compositional ranges greatly expands the half-metallic compositional space towards a set of continuous compact regions (mathematically, each interval in one dimension or a compact region in multiple dimensions is an infinite uncountable set of points).  Properties of half-metals can be continuously tuned within each compositional region.

\acknowledgments This work is supported by the U.S. Department of Energy (DOE), Office of Science, Basic Energy Sciences, Materials Science and Engineering Division.  The research is performed at the Ames Laboratory, which is operated for the U.S. DOE by Iowa State University under contract DE-AC02-07CH11358.

\appendix{}
\section{\label{Appendix}Technical details}

\subsection{\label{subsec:VASP}VASP}
{\par} We use VASP \cite{VASP1,VASP2} 
with the projector augmented waves (PAW) \cite{PAW,PAW2} and Perdew-Wang
(PW91) exchange-correlation functional \cite{PW91} with Vosko-Wilk-Nusair
spin-polarization \cite{VOSKOWN}. 
We use a dense $\Gamma$-centered mesh \cite{MonkhorstPack1976} with at least 72 $k$-points per {\angstrom}$^{-1}$ for the Brillouin zone integration.
The tetrahedron method with Bl\"ochl
corrections \cite{PRB62p6158} is used to calculate electronic density
of states (DOS), while Gaussian smearing with $\sigma=0.05\,$eV is
used within the conjugate gradient algorithm for the full structural
relaxation at zero pressure. 

{\par}We calculate structural energies using either primitive unit cells for ternary line compounds  or supercells for the multi\-component alloys. 
For example,  CuNi$_{3}$Mn$_{4}$Bi$_{4}$ is considered in a 12-atom hH conventional unit cell, while ZnNi$_{7}$Mn$_{8}$Bi$_{8}$
is addressed in a 24-atom $6.141\times6.141\times12.305\,${\angstrom} supercell using $12\times12\times6$ $k$-point mesh.
Co$_2$Mn$_1$(Sn$_n$Sb$_{1-n}$) alloys in Fig.~\ref{Co2MnSnSb_E} are considered using a decorated 16-atom conventional
cubic unit cell of the fH structure, shown in Fig.~\ref{fig1str}. 

{\par}  The width of the band gap is estimated from the plateau in the total number of the minority-spin electrons $n_{s}(E)$, see Fig.~\ref{fig6el} (c, e). 
Counting electrons in the majority-spin channel, we find the number of electrons $n_{-}$ ($n_{+}$), which should be added [$n \! > \! 0$] or removed [$n \! < \! 0$] to move the highest occupied band (the lowest unoccupied band) in the minority spins to $E_{F}$.  These numbers define the compositional interval of half-metallicity in terms of the electron count.

\subsection{\label{subsec:MECCA}Green's Function KKR-CPA }

{\par} The Korringa-Kohn-Rostoker (KKR) \cite{KKR1,KKR2} Green's function code \cite{MECCA} is suitable for both fully ordered and partially disordered multicomponent
systems. It treats multiple scattering in the atomic sphere approximation (ASA) \cite{AJ2009}. 
We use the optimal local basis set \cite{MECCA} with the radii of scattering spheres, determined from the atomic charge distributions around each atom
and its neighbors. 
The periodic boundary correction (Voronoi polyhedra) accounts more properly for both electrostatic energy \cite{TotEnCorr_Christensen1985} and Coulomb potential. 
A variational definition \cite{AJ2012} of the potential-energy zero (so-called muffin-tin zero) \emph{$\text{v}_{0}$} is used and a proper representation of the topology of charge density in the optimal basis set allows to approach accuracy of the full-potential methods \cite{PRB90p205102y2014}.

{\par}Our truncated optimal basis set with $l_{\max}$=3 includes {\emph{s-, p-, d-}} and {\emph{f-}}orbital symmetries; 
we find a negligible sensitivity of energy differences to the higher-order spherical harmonics. 
Integration in a complex energy plane uses the Chebyshev quadrature semicircular contour with 20 points. 
The Brillouin zone integration is performed using a special $k$-points method \cite{MonkhorstPack1976} with a $12\times12\times12$ mesh (and a smaller $8\times8\times8$ supporting grid) for  disordered or ordered systems with 4 atoms per unit cell and $8\times8\times8$ $(6\times6\times6)$
mesh for ordered systems with 16 atoms/cell. 
Homogeneous, substitutional chemical disorder is considered using the coherent potential approximation (CPA) \cite{JohnsonCPA} with the screened-CPA corrections to account for charge-correlations (Friedel screening) associated with local environments \cite{JP1993}. 
Notably, for homogeneous disorder the unit cell remains the same as for the ternary compounds, due to the mean-field CPA configuration averaging simultaneous with the DFT-based self-consistent-field electronic charge.  
Hence, we are able to choose any average composition within the same cell, 
and  to continuously adjust the electron count in direct CPA calculations.

{\par}For exchange-correlation energies and potentials,  we use the generalized gradient approximated (GGA) of 
Perdew, Burke and Ernzerhof revised for solids (PBEsol) \cite{PBESol2008}. 
We perform spin-polarized calculations without spin-orbit coupling. 
Self-consistency is achieved using the modified Broyden's second method \cite{Broyden1988}.

\subsection{\label{subsec:TB-LMTO-vLB}TB-LMTO-vLB}

{\par} We use the corrected exchange $V_{x}+V_{x}^{vLB}$, matched at the ASA radii \cite{VLB, PS2013,PS2015book,PRB93p085204y2016,JPCM2017Singh,PhysRevB.96.054203Singh}, 
in the exchange-correlation energy parameterization \cite{vBH} 
within the local density approximation (LDA).  
An improved ASA basis of TB-LMTO \cite{TBLMTO}
includes atomic spheres and empty spheres (ESs).
The sum of the volumes of all the spheres is equal the total volume of the periodic unit cell. 
The radii of the spheres are chosen to have their overlaps close to the  local  maxima or saddle points  of  the  electrostatic potential.  
The positions of the ES centers are at the high-symmetry points between atoms. 
The relaxed atomic positions are taken from VASP output. 
ESs are treated as empty sites with no cores and small electronic charge density. 
Because the contribution of exchange in empty spheres is very small,
the vLB-correction is calculated only in the atomic spheres  \cite{VLB, PS2013,PS2015book,PRB93p085204y2016,JPCM2017Singh,PhysRevB.96.054203Singh,HSP,HSP1,HSH2014}. 
The core states are treated as atomic-like in a frozen-core approximation. 
The higher-energy valence states are addressed self-consistently in the effective crystal potential. 
The last is the muffin-tin potential. 
The electronic basis combines the plane waves in the nearly free-electron limit outside of the atomic spheres 
and the spherical harmonics inside the atomic spheres, with the matching imposed at the interfaces \cite{TBLMTO}. 
The valence electrons are assumed to be non-relativistic. 

{\par} Self-consistency is achieved when the change of charge density and energy between iterations becomes small: the relative error for averaged charge density is below 10$^{-5}$ and the absolute error in energies is $<10^{-4}$ Ry/atom. 
To facilitate convergence, we use the Anderson mixing. The $k$-space integration is done using the tetrahedron method with the $12\times12\times12$ mesh. 

{\par}
In theory, the Coulomb and exchange self-interactions should be cancelled \cite{PRB93p085204y2016,JPCM2017Singh,PhysRevB.96.054203Singh}. 
The LDA and GGA are known to under-estimate band gaps; this problem is less severe in GGA due to the gradient correction \cite{GGA}. 
On the other hand, the asymptotic vLB correction significantly improves the exchange potential  both near the nucleus  ($r \rightarrow 0$) and at the large-distance limit  ($1/r \to 0$).
It correctly describes the valence and conduction-band energies, and provides the band gaps comparable to those measured in experiments  \cite{PRB93p085204y2016, JPCM2017Singh,PhysRevB.96.054203Singh}.

\subsection{\label{subsec:Mn-form-En}Formation energy of Mn}

To avoid a systematic error in the energy of $\alpha$-Mn, 
resulting from application of DFT to an astonishingly complex $\alpha$-Mn crystal structure with non-collinear moments,  
we use the semi-empirical energy of metallic Mn, 
obtained from the calculated energies of metallic Bi \cite{Complexity11p36y2006} and  
ferrimagnetic MnBi (with NiAs structure) \cite{APLMat2p032103y2014}, 
and the experimental \cite{Shchukarev1961} formation energy of MnBi of $-0.204 \,$eV/f.u.
By construction, the calculated formation energy of
MnBi coincides with this experimental value.

\begin{figure}[ht]
\centering{}\includegraphics[width=75mm]{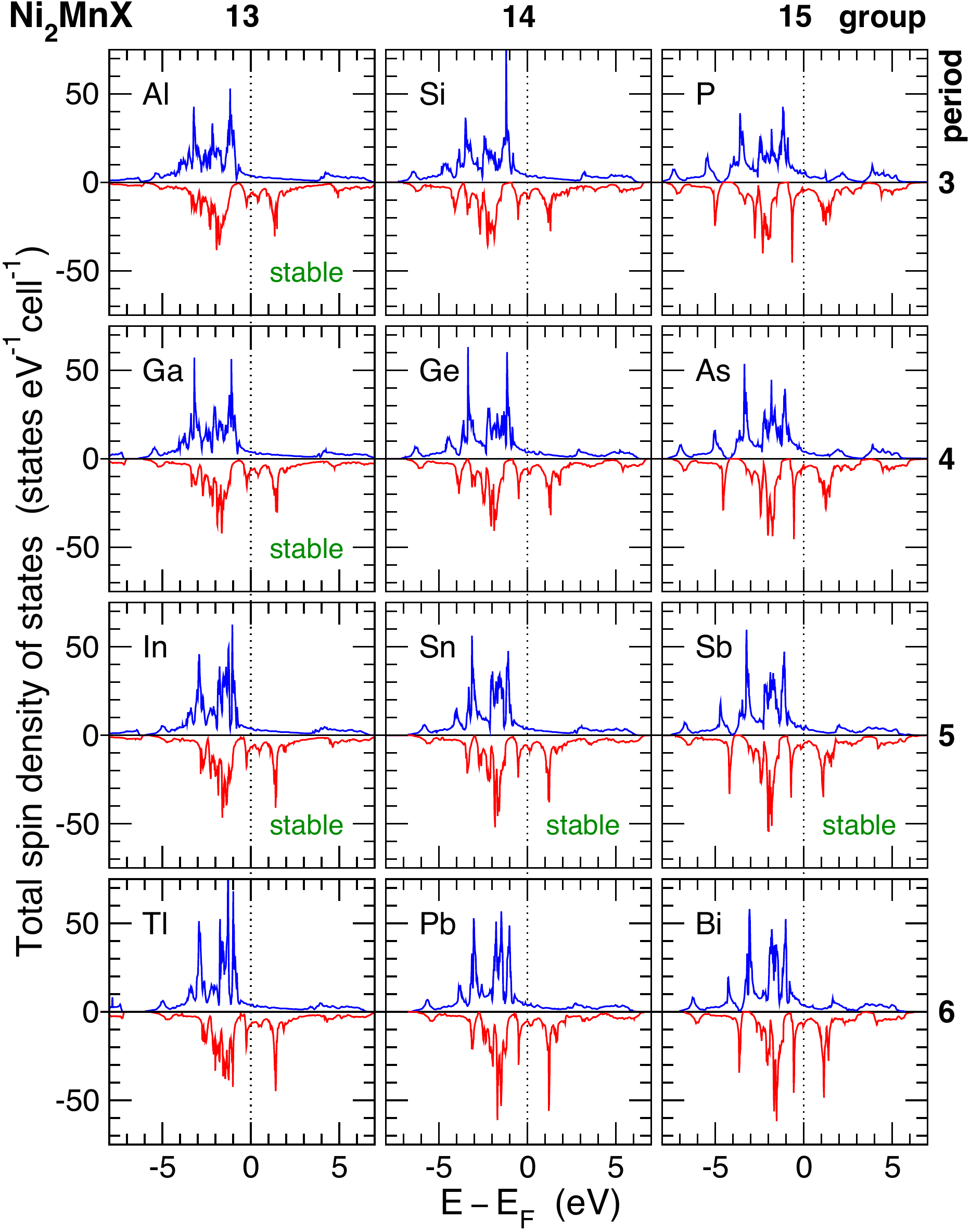} 
\caption{\label{DOS_Ni2MnX_fH} 
Total spin DOS (states/eV$\cdot$f.u.) of fH-Ni$_{2}$MnX for 12 elements X from group 13--15 and periods 3--6). 
All systems are metallic. Compounds that are stable (marked in figure) are known from experiment \cite{JMMM65p76y1987,Uhl1982}.}
\end{figure}

\section{\label{Ni2MnX}  Metallic fH-Ni$_2$MnX} 
{\par} We claim that there are no  half-metals among the fH Ni$_2$MnX alloys. 
We find that the fH Ni$_2$MnX alloys (with X from groups 13--16 and periods 3--6) are metallic, see Fig.~\ref{DOS_Ni2MnX_fH}. 
They have high magnetization (Fig.~\ref{fig2M});  some of them are promising phases for advanced Alnico-type magnets, 
while several might segregate into other compounds (Fig.~\ref{fig4E2}). The electronic-structure calculations reveal  a minimum with a small density of the minority spin states, located $\approx \! 1\,$eV below the Fermi level, see Fig.~\ref{DOS_Ni2MnX_fH}.  We found that there are no half-metals among these alloys,  and we do not see a practical way of transforming them into half-metals by a small amount of doping. 
Thus, we disagree with the suggestion that the rapidly quenched Ni$_{47.3}$Mn$_{28.2}$Sn$_{24.5}$ ribbon, prepared by melt-spinning, was half-metallic \cite{JMMM386p98p2015}.  
We point that although ferromagnets and ferrimagnets can have different conductivity for two spin channels \cite{PRB97p014202y2018}, 
conductivity of half-metals for one of the spins is zero.  We find that Ni$_2$MnSn is magnetic, but not half-metallic.

{\par} The fH~Ni$_{2}$MnX alloys with X=\{P, As, Sb, Bi; S, Se, Te\} are unstable with respect to nickel segregation (see Fig.~\ref{fig4E2}),  because they have a positive formation energy relative to the segregated metallic fcc Ni and hH NiMnX. 

{\par} {\it Mechanical Distortion:} We considered anisotropic distortions of the cubic cells, and found that  fH alloys (including Ni$_{2}$MnSb and Ni$_{2}$MnBi) might distort along the 111 axis with energy lowering, but remain unstable with respect to segregation to fcc Ni and a hH alloy.


\bibliography{Heusler}
\end{document}